\documentclass[12pt]{article}

\usepackage{graphicx}
\usepackage{amsmath}
\usepackage{amssymb}

\def\be{\begin {equation}}

\def\ee{\end {equation}}

\def\bea{\begin{eqnarray}}

\def\eea{\end{eqnarray}}

\begin{document}

\title{ \bf  Resolution of a shock in hyperbolic systems modified by  weak dispersion}

\author{G.A.~El\\
Department of Mathematical Sciences, Loughborough University,\\
Loughborough LE11 3TU, UK
\date{}
}\maketitle

\begin{abstract}
We present a way to deal with dispersion-dominated ``shock-type''
transition in the absence of  completely integrable structure for
the systems that one may characterize as strictly hyperbolic
regularized by a small amount of dispersion.  The  analysis is
performed by assuming that, the dispersive shock transition
between two different constant states can be modelled by an
expansion fan solution of the associated modulation (Whitham)
system for the short-wavelength nonlinear oscillations in the
transition region (the so-called Gurevich -- Pitaevskii problem).
We consider as single-wave  so bi-directional  systems. The main
mathematical assumption is that of hyperbolicity of the Whitham
system for the solutions of our interest. By using general
properties of the Whitham averaging for a certain class of
nonlinear dispersive systems and specific features of the Cauchy
data prescription on characteristics we derive a set of transition
conditions for the dispersive shock, actually bypassing full
integration of the modulation equations.  Along with model KdV and
mKdV examples, we consider a non-integrable system describing
fully nonlinear ion-acoustic waves in collisionless plasma. In all
cases our transition conditions are in complete agreement with
previous analytical and numerical results.

\end{abstract}


\newpage

\section{Introduction}

It is well known  that the resolution of  breaking singularities
in dispersive media occurs through generation of short-wavelength
nonlinear oscillations. The wave-like transition between two
smooth or constant hydrodynamic states is generally called a
dispersive shock (or an undular bore, especially in the context of
water waves). The main observable feature of the dispersive shock
is formation of solitary waves in the vicinity of one of its
edges. At the opposite edge, the wave structure degenerates into
linear wavepacket. The phenomenon of the dispersive shock
formation is quite ubiquitous and its physical contexts range from
gravity water waves and bubbly fluid dynamics to space plasma
physics, fibre optics and Bose-Einstein condensates.

One should be clear from the very beginning that purely dispersive
resolution of a shock is a physical idealization and some amount
of dissipation is inevitably present in real continuous media so
we distinguish between two types of dispersive shocks depending on
the actual role of dissipation relative to that of nonlinearity
and dispersion in the wave development and evolution. There is
some terminological confusion in the literature where the terms
``dispersive shock'',  ``undular bore'' and ``collisionless
shock'' are used as for the dispersion-dominated waves so for the
waves where dispersion and dissipation are in balance. In both
cases, dispersion plays the decisive role in the formation of
local oscillatory structure but the global properties of the
oscillations zone in the two cases are drastically different.

The weakly dissipative undular bores (we use here this term to
distinguish from our main subject -- conservative dispersive
shocks) despite their oscillatory structure exhibit global
properties characteristic for classical, turbulent bores or shock
waves: they have steady (though oscillatory in space) profile and
constant effective width such that the speed of the shock
propagation and the transition conditions can be derived within
the frame of the classical theory of hyperbolic conservation laws
(see \cite{lax73} for instance). The qualitative theory of such
steady undular bores has been first developed by Benjamin and
Lighthill \cite{BL} in the context of shallow-water waves and by
Sagdeev \cite{sagdeev} for rarefied plasma flows. The quantitative
description of the weakly dissipative undular bores has been made
in \cite{johnson}, \cite{GP87}, \cite{AKN87} on the basis of the
unidirectional KdV-Burgers equation and in \cite{EGK05} using the
integrable version of the bi-directional Boussinesq equations
modified by small viscous term.

The developed in \cite{BL} -- \cite{EGK05} theory of steady
undular bores, however, is valid only for the fully established
regime when nonlinearity, dispersion and dissipation are in
balance. Contrastingly, in the {\it dispersion-dominated} case,
the traditional analysis of  the mass, momentum and energy balance
across the undular bore transition can not be applied (at least
directly). The reason for that is that the boundaries of the
dissipationless undular bore (a dispersive shock) diverge with
time, i.e instead of a single shock speed defined by the jump
conditions one has now two different speeds $s^+>s^-$ determining
motion of the transition region boundaries. These speeds ,
however, can not be found without the analysis of the nonlinear
oscillatory structure of the transition region since in the
dissipationless case dispersion not only dramatically modifies the
fine structure of the shock transition but also, along with
nonlinearity, determines its location. The ``conservative''
dynamics of dispersive shocks is of a considerable interest on its
own and also, in many cases,  can be viewed as an unsteady
intermediate asymptotic in a general setting when the small
dissipation is taken into account. Physical examples of such
expanding dispersive shocks include atmospheric undular bores
(morning glory) \cite{PS}, optical shocks  in  the long-distance
optical communication systems \cite{kod99} and collisionless
shocks in Bose-Einstein condensates  \cite{bec}.

In  a weakly nonlinear case, when  the original conservative
system can be approximated by one of  the exactly integrable
equations, the study of the  dispersive shock phenomenon have
stimulated discovery of a whole new class of mathematical problems
which can be broadly described as the singular semi-classical
limits in integrable systems. In the Lax-Levermore-Venakides
theory \cite{laxlev83}, \cite{ven85} (see also  \cite{llv93})
developed originally for the KdV equation, and more recently
extended to the defocusing NLS \cite{lev99} and focusing mKdV
\cite{lev03} equations
 the evolution of the dispersive shock is modelled by the
zero dispersion
 limit of the exact multi-soliton (multi-gap)
solution of the original dispersive wave equation. The main
characteristic feature of the zero-dispersion limit is
co-existence of smooth and rapidly oscillating regions in the
solution after the breaktime. In the smooth regions the
zero-dispersion limit exists in a strong sense and is given by the
classical solution of the dispersionless equation whereas in the
oscillating regions this limit exists in a weak, averaged sense
and is governed by a certain system of quasi-linear equations {\it
different from the dispersionless limit}. This system turned out
to coincide with the modulation equations obtained in 1960's by
Whitham \cite{wh65} by averaging nonlinear single-phase
wavepackets and later generalised by Flaschka, Forest and
McLaughlin \cite{ffm79} to a multiphase case. The subsequent
development of the Lax-Levermore theory \cite{ven90}, \cite{llv93}
has shown that the local waveform in the oscillating regions is
indeed described by the multiphase solutions. The weak limits in
the Lax-Levermore problem then can be regarded as the result of
the Whitham averaging over these solutions.

A direct formulation of the dispersive shock problem in terms of
the Whitham equations had been proposed (although without rigorous
justification) much earlier by Gurevich and Pi\-ta\-ev\-skii (GP)
\cite{gp74} who supplied the Whitham system for the KdV equation
with natural matching conditions at the boundaries separating
smooth and oscillating regions and solved the problem analytically
for the initial data in the form of a step. The discovery by
Tsarev of the generalised hodograph method \cite{ts85}, and
Krichever's algebro-geometric construction \cite{kr88} allowed for
obtaining of many new exact global solutions to the Whitham
equations (see for instance \cite{ekv01}, \cite{gr04} and
references therein). Availability of such solutions heavily relies
on the integrability of the Whitham systems considered and, in
particular, on the existence of the Riemann invariant
representation.

Thus, either way, integrability seems to be an essential part of
the analytic theory of dispersive shocks. However, in many
physically relevant situations, integrable systems although
providing valuable insight into the qualitative properties of
nonlinear dispersive wave propagation fail to yield satisfactory
quantitative agreement with experimental data (see for instance
discussion in \cite{cc99} in the context of finite-amplitude
internal shallow water waves). This explains the growing interest
in the derivation of relatively simple non-integrable models
which, while providing a more accurate description of physical
effects, still seem to be amenable to analytic treatment. In the
context of the dispersive shock theory this interest is supported
by a strong numerical evidence that such features of
zero-dispersion limits of integrable systems as formation of
oscillatory zones and weak convergence are true for the systems
that are not completely integrable but are structurally similar to
integrable ones (see for instance \cite{llv93} \cite{hl91},
\cite{gm84}). On the other hand, the Whitham method used in the
direct Gurevich-Pitaevskii description of dispersive shocks does
not require integrability from the governing system. What is
needed is just the existence of the periodic travelling wave
solutions and availability of sufficient number of conservation
laws. So, in the absence of rigorous general approach it seems a
natural idea to {\it assume} the single-phase Whitham description
for a ``non-integrable'' dispersive shock and explore analytic
consequences of such an assumption. Hyperbolicity of the Whitham
equations (which also has to be assumed, for instance, on the
grounds of numerical evidence of modulational stability) would be
an essential part of such a problem formulation. Of course, such a
heuristic approach requires validation. This can be made by (i)
testing its results on integrable systems where exact solutions
are available and (ii) by comparison  with available numerical
results for non-integrable systems.

In this paper, we develop the outlined approach for the the decay
of an initial step problem for a broad class of  systems that one
may characterise as strictly hyperbolic modified by weak
dispersion. It is clear that in the step decomposition problem the
solution of the (quasilinear) Whitham equations must depend on
$x/t$ alone, which implies principal availability of several
integrals of motion for their self-similar reductions. It is also
clear that existence of the similarity solutions does not rely on
complete integrability or existence of the Riemann invariants for
the Whitham equations ( this, unfortunately, does not guarantee
that such a solution would be available analytically in the
absence of the Riemann invariants). However, the Whitham equations
being obtained by averaging over periodic family have a number of
special properties distinguishing them from the general class of
hyperbolic quasi-linear systems. In particular, they allow for
exact reductions to the original dispersionless equations in the
zero-amplitude (linear) and zero-wavelength (soliton) limits. This
general property together with the self-similarity imposes a
number of restrictions on possible values of the modulation
parameters at the linear and soliton edges of the dispersive shock
defined by natural boundary conditions. We derive these
restrictions by finding a set of integrals along the edge
characteristics of the Whitham expansion fan. These integrals, of
course, are equivalent to the basic integrals for the similarity
solution evaluated at the edge points. A surprising fact is that,
the edge parameters are in a simple and universal way expressed in
terms of the linear dispersion relation of the system and one of
its nonlinear ``dispersionless'' characteristic velocities.

The obtained set of restrictions can be viewed as a ``dispersive''
replacement for the classical shock conditions and includes for a
general bi-directional case

(i) a condition for admissible jumps for hydrodynamic variables
across the dispersive shock (which {\it does not} coincide with
the classical jump condition);

(ii) the speeds of the dispersive shock edges as functions of a
given jump satisfying (i);

\noindent and, since the parameters in (ii) turn out to be
determined not uniquely,

 (iii) the set of inequalities selecting the unique valid set of
 parameters (ii) and providing consistency of the whole
 construction. These inequalities represent an analog of entropy
 conditions known in  classical gas dynamics.

Thus, the obtained conditions allow one to derive main
quantitative characteristics of the dispersive shock transition
bypassing the integration of the Whitham system. In particular,
they allow for obtaining the amplitude of the leading (or trailing
depending on the sign of the dispersion in the system) solitary
wave in the dispersive shock, the major  parameter in
observational/experimental  data. For instance, the conditions
readily yield the well-known result of the original
Gurevich-Pitaevskii paper \cite{gp74} which reads that the
amplitude of the lead solitary wave in the KdV dispersive shock is
$2 \Delta$ where $\Delta$ is the value of the initial step.

We apply the obtained general dispersive shock  conditions to two
integrable (KdV and defocusing mKdV) equations and one
non-integrable system describing fully nonlinear ion-acoustic
waves in collisionless plasma. In all cases our description is in
complete agreement with previous analytic/numerical results.  We
also discuss the accuracy and  some restrictions of the developed
approach.

\section{Gurevich-Pitaevskii problem for the KdV equation}
 We start with an exposition of the  theory of Gurevich
and Pitaevskii (GP) \cite{gp74} formulated originally for the KdV
equation and later generalised to other integrable equations by
different authors (see \cite{kam00} for a detailed introduction
and many useful references).  Although all formulas in this
section are
 known very well, it is instructive for the purposes
of this paper to have them handy as they will allow us later to
draw parallels with  ``non-integrable'' theory.

We take the KdV equation  in the form
\begin{equation}\label{kdv}
    u_t+uu_x+u_{xxx}=0\, .
\end{equation}
Let the initial perturbation $u(x,0)=u_0(x)$ have the form of a
large-scale ($\Delta x \gg 1$) monotonically decreasing function
with a single inflection point. As a special important case, one
considers a smooth step
\begin{equation}\label{disc}
u_0(-\infty) = u^- \, , \quad  u_0(+\infty) = u^+ \, , \quad
u_0'(x) <0 \, , \, \
\end{equation}
so that the characteristic width of the transition region, say $l$
is finite. Of course, without loss of generality the initial step
for the KdV equation can be normalized to unity but for our
purposes it is convenient to retain arbitrary values for $u^-,
u^+$; we only assume $u^->u^+$.

The qualitative picture of the KdV evolution of a smooth monotonic
profile is as follows. During  the initial stage of evolution,
$|u_{xxx}|\ll |uu_x|$ and one can neglect the dispersive term in
the KdV equation (\ref{kdv}). The evolution  at this stage is
approximately described by the dispersionless (classical) limit of
the KdV equation:
\begin{equation}\label{hopf}
u \approx \beta(x,t): \quad \beta_t+\beta\beta_x=0\, , \quad
\beta(x,0)=u_0(x)\,  .
\end{equation}
 The evolution (\ref{hopf}) leads to a gradient catastrophe :
$t \to t_{b}: \quad \beta_x \to -\infty$. For $t > t_b$ (without
loss of generality we put $t_b=0$), the dispersive term $u_{xxx}$
should be taken into account in the vicinity of the breaking point
and, as a result, the regularization of the singularity happens
through the generation of small-scale nonlinear oscillations
confined to a finite, albeit expanding, space region. This
oscillatory structure represents a dispersive analog of a shock
wave. Near the leading edge of such a {\it dispersive shock} (or
an {\it undular bore} in a different terminology) the oscillations
are close to successive solitary waves and in the vicinity of the
trailing edge they are nearly linear (see Fig.~1). This structure
has been recovered in a rigorous theory using the complete
integrability of the KdV equation (see for instance \cite{dvz94})
but in the Gurevich-Pitaevskii approach it is an assumption based,
say,  on the results of numerical simulations. This approach is
consistent with the aims of this paper which develops a way to
deal  with generally non-integrable systems, where the IST
formalism is not available in principle.
\begin{figure}[ht]
\centerline{\includegraphics[width=8cm,height=
5cm,clip]{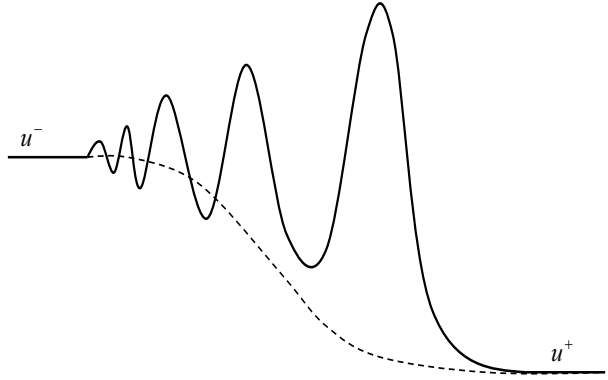}} \vspace{0.3 true cm} \caption{Oscillatory
structure of the dispersive shock evolving from the initial step
(dashed line).} \label{fig1}
\end{figure}

We shall model the local waveform of the dispersive shock  by the
single-phase periodic solution of the KdV equation travelling with
constant velocity $c$: $u(x,t)=u(\theta)$, $\theta=x-ct$. This
solution is specified by the ordinary differential equation
\begin{equation}\label{trcn}
(u_\theta)^2= -G(u) \, ,   \qquad u(\theta + 2\pi/k)=u(\theta)\, ,
\end{equation}
where
\begin{equation}\label{Q}
 G(u)= \frac{1}{3}(u-u_1)(u-u_2)(u-u_3)\, ,
\end{equation}
$u_3 \ge u_2 \ge u_1$ being constants of integration. The phase
velocity $c$ and the wavenumber $k$ are expressed in terms of the
roots $u_j$ as
\begin{equation}\label{L}
c=\frac{1}{3}(u_1+u_2+u_3)\, , \quad k={\pi}
\left(\int\limits^{u_3}_{u_2}\frac{du}{\sqrt{-G(u)}}\right)^{-1}=\frac{\pi}{2\sqrt{3}}
\frac{(u_3-u_1)^{1/2}} { K(m)}\, ,
\end{equation}
where $K(m)$ is the complete elliptic integral of the first kind.
The modulus $m$ and the amplitude $a$ of the travelling wave are
expressed in terms of $\{u_j\}$ as
\begin{equation}\label{m}
 m=\frac{u_3-u_2}{u_3-u_1}\, \qquad a=u_3-u_2\, .
\end{equation}
Eq. (\ref{trcn}) is integrated in terms of Jacobian elliptic
function $\hbox{cn}(\xi;m)$ to give, up to an arbitrary phase
shift,
\begin{equation}\label{cn}
u(x,t)= u_2+ a \,
\hbox{cn}^2\left(\sqrt{\frac{u_3-u_1}{3}}(x-ct);m\right)\, .
\end{equation}
 When $m \to 0$ $(u_2 \to u_3)$ the solution (\ref{cn})
turns into the vanishing amplitude harmonic wave
\begin{equation}\label{lin0}
u(x,t) \approx u_3-a\sin^2[k_0(x-c_0t)]\,, \quad a=u_3-u_2 \ll 1
\, ,
\end{equation}
where $k_0=k(u_1, u_3,u_3)$, $c_0=c(u_1, u_3, u_3)$. The
relationship between $c_0$ and $k_0$ is obtained from
Eqs.~(\ref{L}) considered in the limit $u_2 \to u_3$:
\begin{equation}\label{dr0}
c_0=u_3-k_0^2 \,
\end{equation}
and agrees with the KdV linear dispersion relation for
small-amplitude waves propagating on the background $u=u_3$.

When $m \to 1$ $(u_2 \to u_1)$, the cnoidal wave (\ref{cn}) turns
into a soliton
\begin{equation}\label{sol00}
u_s(x,t)=u_1+ a_s \hbox{sech}^2[\sqrt{a_s/3}(x-c_st)]\, ,
\end{equation}
which speed $c_s=c(u_1, u_1, u_3)$ is connected with the amplitude
$a_s$ by
\begin{equation}\label{sdr0}
c_s=u_1+a_s/3 \, .
\end{equation}
In Eqs.~(\ref{sol00}), (\ref{sdr0}), $u_1$ plays the role of a
background which can be put equal to zero for the uniform
solutions (as well as $u_3$ in (\ref{lin0}), (\ref{dr0})).
However, this is not the case if one is interested in modulated
travelling waves where the local integrals $u_j$ become slow
functions of $x$ and $t$ and the background term can not be
eliminated by the passage to the moving reference frame. This
simple fact will play an important role in what follows.

We now consider slowly modulated cnoidal waves by letting the
constants of integration $u_j$ be functions of $x$ and $t$ on a
large spatio-temporal scale $\Delta x  \gg 1$, $\Delta t \gg 1$.
Then the evolution  equations for $u_j(x,t)$ can be obtained by
averaging  any three  KdV conservation laws $\partial_t P_j
+\partial_x Q_j=0$ over the period of the travelling wave
(\ref{cn}) \cite{wh65} and are referred to as the modulation, or
Whitham, equations. The averaging is made according to
(\ref{trcn}) as
\begin{equation} \label{av}
  \overline{F}(u_1,u_2,u_3)= \frac{k}{2\pi}\int \limits _0 ^{2\pi/k}F(u(\theta;u_1,u_2,u_3))d\theta=
  \frac{k}{\pi}\int
\limits^{u_3}_{u_2}\frac{F(u)}{\sqrt{-G(u)}}du \,
 .
\end{equation}
In particular, the mean  is calculated as
\begin{equation}\label{mean}
\bar u= u_1+2(u_3-u_1)E(m)/K(m)\, ,
\end{equation}
where $E(m)$ is the complete elliptic integral of the second kind.

As a result, the KdV modulation system is  obtained in a
conservative form
 \begin{equation}\label{avkdv}
\frac{\partial } {\partial t}\overline{P}_j(u_1,u_2, u_3)
+\frac{\partial }{\partial x}\overline{Q}_j(u_1, u_2, u_3)=0\, ,
\quad j=1,2,3 \, ,
\end{equation}
where $\bar P_j, \bar Q_j$ are expressed in terms of the complete
elliptic integrals.

One of the modulation equations can be replaced by the so-called
wave number conservation law, which represents a consistency
condition in the formal perturbation procedure equivalent to the
Whitham averaging (see e.g. \cite{dn89}),
\begin{equation}\label{wc0}
\frac{\partial k}{\partial t} + \frac{\partial \omega}{\partial
x}=0\, ,
\end{equation}
where $\omega=kc$  is the frequency. Of course, Eq.~(\ref{wc0})
can be obtained as a consequence of the three modulation equations
(\ref{avkdv}) \cite{wh65}.

It has been discovered in \cite{wh65} that, upon introducing
symmetric combinations
\begin{equation}\label{ri}
\beta_1=\frac{u_1+u_2}{2}\, , \ \beta_2=\frac{u_1+u_3}{2}\, , \
\beta_3=\frac{u_2+u_3}{2}\,
\end{equation}
the system (\ref{avkdv}) reduces to its diagonal (Riemann) form
\begin{equation}\label{rkdv}
\frac{\partial \beta_j}{\partial t}+V_j(\beta_1, \beta_2,
\beta_3)\frac{\partial \beta_j}{\partial x}=0 \, , \quad
j=1,2,3\,.
\end{equation}
where the characteristic velocities $V_3 \ge V_2 \ge V_1$  are
certain combinations of complete elliptic integrals of the first
and the second kind. They can be conveniently represented in a
``potential" form \cite{gke91}, \cite{kud91}
\begin{equation}\label{vj}
V_j=\frac{\partial (kc)}{\partial\beta_j}/\frac{\partial
k}{\partial\beta_j}\,
\end{equation}
following from the consistency of wave number conservation law
(\ref{wc0}) with the Riemann system (\ref{rkdv}). Here $k$, $m$
and $c$ are expressed in terms of the Riemann invariants as
\begin{equation}\label{kb}
k=\frac{\pi}{\sqrt 6}\frac{(\beta_3-\beta_1)^{1/2}}{ K(m)}\, ,
\quad m=\frac{\beta_2-\beta_1}{\beta_3-\beta_1}\, \quad
c=\frac{1}{3}(\beta_1 + \beta_2 +\beta_3).
\end{equation}
Of course, the  existence of the Riemann invariants for the
quasi-linear system of the third order (\ref{avkdv}) is a
nontrivial fact and this feature is connected with the
preservation of the KdV integrability under the averaging. A
regular way of obtaining the KdV-Whitham system in the Riemann
form  has been developed by Flaschka, Forest and McLaughlin
\cite{ffm79} using the methods of finite-gap integration. For the
single-phase case of our interest an elementary general approach
not requiring algebraic geometry can be found in the monograph
\cite{kam00}. The explicit expressions for $V_j$ in terms of the
complete elliptic integrals of the first and the second kind can
be found elsewhere (see for instance \cite{ts80}, \cite{dn89}). We
will present here only their limiting properties.

In the harmonic limit
\begin{equation}\label{hv}
m=0: \qquad V_3 = \beta_3\, , \quad  V_2=V_1= 2\beta_1-\beta_3
\end{equation}
so that the Whitham system reduces to
\begin{equation}\label{wh0}
\beta_2=\beta_1\, , \quad \frac{\partial \beta_3}{\partial t}+
\beta_3 \frac{\partial \beta_3}{\partial x}=0\, ,  \quad
 \frac{\partial \beta_1}{\partial t}+(2\beta_1-\beta_3)
 \frac{\partial \beta_1}{\partial
 x}=0 \, .
\end{equation}
 In the soliton limit
\begin{equation}\label{sv}
m=1: \qquad V_2=V_3= \frac{1}{3} (\beta_1+2\beta_3)\, , \quad
V_1=\beta_1
\end{equation}
and the Whitham system reduces to
\begin{equation}\label{whs}
\beta_2=\beta_3\, , \quad \frac{\partial \beta_1}{\partial t}+
\beta_1 \frac{\partial \beta_1}{\partial x}=0\, ,  \quad
 \frac{\partial \beta_3}{\partial t}+ \frac{1}{3}(\beta_1+2\beta_3)\frac{\partial \beta_3}{\partial
 x}=0 \, .
\end{equation}
Thus, the Whitham system admits nontrivial {\it exact} reductions
to the dispersionless limit via singular limiting transitions
$\beta_2 \to \beta_1$ (linear limit) and $\beta_2 \to \beta_3$
(soliton limit). In both limits one of the Whitham equations
converts into the Hopf equation, while the remaining two  merge
into one for the Riemann invariant along a double characteristics.
Such a special structure of the Whitham equations makes it
possible to formulate the following natural matching problem for
(\ref{rkdv}) \cite{gp74}.

Let the upper $(x,t)$ half-plane  be split into three domains :
$\{(x,t>0): \ \ (-\infty, x^-(t)) \cup [x^-(t), x^+(t)] \cup
(x^+(t), +\infty)\}$ (see Fig. 2), in which the solution is
governed by different equations: outside the interval $[x^-(t),
x^+(t)]$ it is governed by the dispersionless limit (\ref{hopf})
while within the interval $[x^-(t), x^+(t)]$ the dynamics is
described by the Whitham equations (\ref{rkdv}) so that the
following matching conditions must be satisfied:
\begin{equation}
\begin{array}{l}
x=x^-(t):\qquad \beta_2=\beta_1\, , \ \ \beta_3 = \beta \\
x=x^+(t):\qquad \beta_2=\beta_3\, , \ \ \beta_1 = \beta
\end{array}
\label{bc}
\end{equation}
where $\beta(x,t)$ is the solution of the Hopf equation
(\ref{hopf}) and the (free) boundaries $x^{\pm}(t)$ are unknown at
the onset.

\begin{figure}[ht]
\centerline{\includegraphics[width=10cm,height=
7cm,clip]{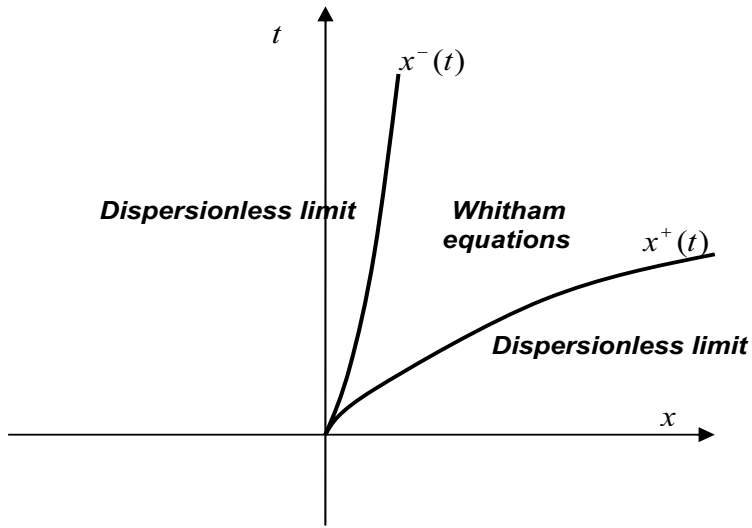}} \vspace{0.3 true cm} \caption{ Splitting of
the $(x,t)$ - plane in the Gurevich -- Pitaevskii problem}
\label{fig2}
\end{figure}

{\bf Remark.} One can notice that  formulated in this way, the GP
problem contains an implicit assumption about the spatial
structure of the dispersive shock, namely, it identifies the
leading edge with the soliton and the trailing edge with the
linear wave from the very beginning. Of course, for the KdV
equation this wave pattern can be inferred from  previous
numerical simulations or from simple physical reasoning that owing
to the  negative sign of the dispersion in the KdV equation, the
longer waves must propagate with greater speed. However, from
mathematical point of view, such an assumption must be confirmed
or rejected by the actual solution so, strictly speaking, at this
point we actually don't know if $x^+>x^-$.

For the initial data with a single breaking point, the conditions
(\ref{bc}) uniquely define the global solution $\beta_j(x,t)$ of
the Whitham equations (\ref{rkdv}) (see \cite{dn89}). It follows
from Eq.~(\ref{bc}) and the limiting properties of the Whitham
velocities (\ref{hv}), (\ref{sv}) that the free boundaries
$x^{\pm}(t)$ are determined by the double eigenvalues of the
Whitham system for $m=0$ ($x=x^-(t)$) and $m=1$ ($x=x^+(t)$) and
are found from the ordinary differential equations
\begin{equation}\label{b0}
 dx^-/dt=V_{2}(\beta_1, \beta_1, \beta_3)|_{x=x^-}= V_{1}(\beta_1, \beta_1, \beta_3)|_{x=x^-}\equiv
 V^-(x^-, t)\, ,
 \end{equation}
\begin{equation}\label{bs}
dx^+/dt=V_{2}(\beta_1, \beta_3, \beta_3)|_{x=x^+}= V_{3}(\beta_1,
\beta_3, \beta_3)|_{x=x^+} \equiv V^+(x^+, t)
\end{equation}
where $\{\beta_j=\beta_j(x,t)\}$ is the  solution of the GP
problem. We note that the curves $x^{\pm}(t)$ are sometimes
referred to as the phase transition boundaries since they separate
the regions of the zero-phase (external smooth flow) and the
single phase (dispersive shock) solutions.

The local integrability of the Riemann system (\ref{rkdv}) is
based on certain relationships between the characteristic
velocities and has been established by Tsarev  \cite{ts85} who
proposed a generalisation of the classical hodograph method
applicable to the hydrodynamic-type systems with the number of
field variables exceeding two.  There is, however, a special
important case when the generalised hodograph transform
degenerates and the Tsarev integrability scheme becomes redundant.
This is the case of the decay of an initial step (\ref{disc}),
which implies that the modulation variables are the functions of
$s=x/t$ alone for $t>>l$, where $l$ is the characteristic width of
the (smooth) initial step.  In this case, the GP problem has the
asymptotic solution in the form of the centred simple wave in
which all but one Riemann invariants are constant:
\begin{equation}\label{simkdv}
\beta_1=u^+\, , \ \ \beta_3=u^-\, , \ \ V_2(u^-, \beta_2, u^+)=s
\, ,
\end{equation}
or explicitly,
\begin{equation}\label{vs0}
\frac{1}{3}(u^-+2u^++m\Delta)  - \frac{2}{3}
\frac{(1-m)m\Delta}{E(m)/K(m)-(1-m)}=\frac{x}{t}\,,
\end{equation}
where $\Delta=u^--u^+$ is the magnitude of the initial jump. Since
the solution (\ref{vs0}) represents a characteristic fan it never
breaks for $t>0$ and therefore is global. The self-similar
boundaries $s^{-}$ and $s^{+}$ of the dispersive shock are found
from the solution (\ref{vs0}) by putting in it $m=0$  and $m=1$
 respectively:
\begin{equation}\label{spm0}
s^-=s(0)=u^+ - \Delta \, , \quad s^+=s(1)=u^++\frac{2}{3}\Delta \,
,
\end{equation}
One can see that $s^+>s^-$  so the assumed wave pattern with the
leading soliton and trailing linear wavepacket was indeed correct.
Now  the amplitude of the lead soliton is simply
\begin{equation}\label{as0}
a^+=2(\beta_3-\beta_1)=2\Delta \, ,
\end{equation}
whereas the value of the wavenumber of the trailing wave packet
follows from Eq.~(\ref{kb}a) evaluated for $\beta_2=\beta_1$ which
yields
\begin{equation}\label{k0}
k^-=\sqrt{\frac{2}{3}(\beta_3-\beta_1)}=\sqrt{\frac{2}{3}\Delta}
\, .
\end{equation}
We see that the simple form (\ref{simkdv}) of the analytic
solution to the GP problem is possible owing to the availability
of the Riemann invariant form (\ref{rkdv}) which admits
$\beta_j=constant$ as an exact solution. The formulas (\ref{spm0})
-- (\ref{k0}) are obtained as consequences of this  global
solution. So, since the existence of the Riemann invariants is due
to complete integrability of the original KdV equation one may
conclude that the possibility of obtaining the ``global'' formulas
(\ref{spm0})--(\ref{k0}) relies on the integrability as well. This
is  not in fact so. In the next section, we will show that these
formulas can be drawn directly from the Whitham system in
``physical" variables using some very general properties of the
Whitham systems subjected to natural boundary (matching)
conditions of the GP type. Actually, Eqs.~(\ref{as0}), (\ref{k0})
will be obtained without making use of the Riemann invariant form
for the Whitham equations, and bypassing their explicit
integration.
\section{``Non-integrable'' reformulation of the  Gurevich - Pitaevskii problem}
\subsection{Formulation}
Let us reformulate the GP problem in terms of the Whitham system
in its original, non-diagonal form (\ref{avkdv}). Along with the
set of natural ``mathematical" variables $u_1, u_2, u_3$ we will
use an equivalent set of ``physical" variables $ \bar u, k, a$
which are connected with $u_1, u_2, u_3$ by means of
Eqs.~(\ref{mean}), (\ref{L}), (\ref{m}).

From here on we are not going to exploit  subtle algebraic
properties underlying the integrability of (\ref{avkdv}). Instead,
we will use some general properties of the system (\ref{avkdv})
connected with its "averaged" origin and distinguishing it from a
general class of hyperbolic quasi-linear systems of the third
order. In all relevant cases we will indicate if the obtained
relationships could be extended to a more general  context.

We start with the KdV modulation system in conservative form (see
\cite{wh65} for instance)
\begin{equation}\label{wh2}
\frac{\partial \bar u}{\partial t}+  \frac{\partial }{\partial
x}\left(\frac{\overline{u^2}}{2}\right)=0 \, , \quad
\frac{\partial}{\partial t}\left( \frac{\overline{u^2}}{2}\right)+
\frac{\partial }{\partial x}\left(\overline {\frac{u^3}{3}
-\frac{u_\theta^2}{2} + uu_{\theta \theta}} \right)=0 \, , \quad
\frac{\partial k}{\partial t} + \frac{\partial \omega (\bar u, k,
a)}{\partial x}=0\, .
\end{equation}
It is not difficult to show that the  averaging (\ref{av}) over
the period of the travelling wave specified by Eqs. (\ref{trcn}),
(\ref{Q}) implies the following general relationships for the
averaged variables in the harmonic ($m=0$) and soliton ($m=1$)
limits
\begin{equation}\label{31}
\overline{F(u)}|_{u_2 = u_3}=F(u_3);  \qquad \overline{F(u)}|_{u_2
= u_1}=F(u_1);
\end{equation}
In particular,  this implies that $\bar u (u_1, u_3,u_3)= u_3$ and
$\bar u (u_1,u_1, u_3)=u_1$ (this can also be directly seen from
the explicit expression (\ref{mean})). Then, since $u_2= u_3$ is
equivalent to $a= 0$ and $u_2 = u_1$
 to $k = 0$, the
relationships (\ref{31}) assume in the physical variables the form
\begin{equation}\label{311}
\overline{F(u)}|_{a = 0}=F(\bar u); \qquad \overline{F(u)}|_{k =
0}=F(\bar u);
\end{equation}
The relationships (\ref{311}) immediately imply that in both
harmonic and soliton limits, $\overline{u^2}=\bar{u}^2$ hence the
first modulation equation (\ref{wh2}) turns into the Hopf equation
for the mean value $\bar{u}_t+\bar{u}\bar{u}_x=0$. It can also be
readily show shown using (\ref{trcn}), (\ref{av}) that the same is
true for the second modulation equation (\ref{wh2}) as well.
Therefore:

i) the modulation system (\ref{wh2}) admits {\it exact} reduction
to a lower order system both for $a=0$ and $k=0$. The limiting
transitions $a \to 0$ and $k \to 0$ are, therefore, singular;

ii) the limiting transitions $a \to 0$ and $k \to 0$ provide two
possible ways of passage to the dispersionless limit {\it in the
modulation equations}.

We note that the same conclusions have been inferred from the
Riemann equations (\ref{wh0}), (\ref{whs}).

Of course, the described limiting behaviour is quite expected
 for a modulation system since in both afore-mentioned limits the
oscillations do not contribute into the averaging hence the
averaged (semi-classical) system must agree with non-oscillatory
classical limit  of the original equation. It is also clear  that
this fact is not  unique to the KdV equation and must hold for
other dispersive-hydrodynamics systems supporting the periodic
travelling waves with the quadratic behaviour of the potential
curve $G(u)$ (\ref{Q}) near its extrema. One can see, in
particular, that this (generic for weakly dispersive systems)
behaviour of the potential curve guarantees exponential decay for
the soliton solutions so that they do not contribute into the
averaging.

Now the GP natural boundary conditions (\ref{bc}) can be
reformulated in terms of the matching of the mean flow in the
oscillatory region with the smooth external flow
\begin{equation}
\begin{array}{l}
x=x^-(t):\qquad a=0\, , \ \ \bar u = \beta\, , \\
x=x^+(t):\qquad k=0\, , \ \ \bar u = \beta \, ,
\end{array}
\label{bc2}
\end{equation}
where $\beta(x,t)$ is defined by the classical limit (\ref{hopf}).
Using explicit expression for $\bar u$ (\ref{mean}) and the
relationships (\ref{ri}) one can easily see that the conditions
(\ref{bc2}) indeed equivalent to (\ref{bc}). For the particular
case of the decay of an initial discontinuity (\ref{disc}), the
conditions (\ref{bc2}) assume the form
\begin{equation}
\begin{array}{l}
x=s^-t : \qquad a=0\, , \ \ \bar u =u^- \, , \\
x=s^+t : \qquad k=0\, , \ \ \bar u =u^+ \, .
\end{array}
\label{bc3}
\end{equation}
We note  that  when passing from (\ref{bc2}) to (\ref{bc3}) we
have made an implicit assumption about the structure of the
dispersive shock by implying that $s^+>s^-$.

 The boundaries $x^{\pm}(t)$ of the modulation solution are
defined by Eqs.~(\ref{b0}), (\ref{bs}) in terms of the
characteristic velocities of the modulation systems where either
$m=0$ (trailing edge) or $m=1$ (leading edge). This definition, of
course, does not rely on the existence of the Riemann invariants
and must hold in  general case. This can be explained in the
following way. It is clear that in order to provide continuous
matching of the solutions of two (consistent) quasilinear
hyperbolic systems of different order (the Whitham system and the
Hopf equation in our case) the matching lines must {\it
necessarily} be the multiple characteristics for the system with a
higher order. Thus, the natural matching conditions (\ref{bc2})
must be supplemented with the definition of the boundaries
$x^{\pm}(t)$ in terms of the double eigenvalues of the modulation
system (see \cite{te99} for a detailed description of the
characteristics behaviour in the GP problem).

We represent the Whitham system (\ref{avkdv}) in a generic
quasi-linear form
\begin{equation}\label{ql}
{\bf y}_t+A({\bf y}){\bf y}_x=0 \, ,
\end{equation}
where ${\bf y}=(\bar u, k, a)^T$ and $A({\bf y})$ is the
coefficient matrix.
 As is well known (see for instance
\cite{Wh74}, \cite{ry78}) the  quasilinear system (\ref{ql}) can
be represented in a characteristic form
\begin{equation}\label{char}
\sum _{j=1}^3 {b^{(m)}_j}\frac{d_m y^j}{dt}=0\, , \quad
\frac{d_m}{dt} \equiv  \frac{\partial }{\partial t }+V_m
\frac{\partial }{\partial x}\, , \quad m=1,2,3 \, ,
\end{equation}
where the characteristic velocities $V_j(\bf{y})$  are the
eigenvalues  of the matrix $ A({\bf y})$, and ${\bf b^{(m)}}({\bf
y})$ is its left eigenvector corresponding to $m$-th eigenvalue:
${\bf b^{(m)}}A=V_m{\bf b^{(m)}}$. We assume that the
characteristic velocities $V_1({\bf y}) \le V_2({\bf y}) \le
V_3({\bf y})$ are real for the solutions of our interest so the
system (\ref{ql}) is hyperbolic (of course, for the KdV equation,
the hyperbolicity of the Whitham system is a proven fact
\cite{Lev} but in our, ``non-integrable" approach it is an
assumption).

It can be readily seen using an equivalent characteristic
representation (\ref{char}) in ``mathematical'' coordinates, ${\bf
y} \mapsto {\bf y^*}=(u_1,u_2,u_3)$, that the fact that the
modulation system (\ref{ql}) admits exact reductions for
$u_2=u_{3}$ ($a=0$) and $u_2=u_1$ ($k=0$)implies that the multiple
roots of the potential curve $G(u)$ necessarily correspond to
multiple eigenvalues of the averaged system. This, again, is a
consequence of the properties of the Whitham averaging (\ref{av}).
It follows from the characteristic velocities ordering that the
double eigenvalues can be either $V_2=V_{3}$ or $V_2=V_1$. The
remaining single eigenvalue in both cases is $\bar u$ since the
Hopf equation is an exact reduction of the Whitham system in both
limits. If the explicit expressions for the characteristic
velocities are known, the correspondence between the double roots
of the travelling wave potential function $G(u)$ and the double
eigenvalues of the averaged system is established directly (see
(\ref{hv}), (\ref{sv})). Alternatively, one can perform an
asymptotic analysis of the modulation system for $m \ll 1$ ($a \ll
1)$ and $(1-m) \ll 1$ ($k \ll 1$) and establish the sought
correspondence. Either way, even without additional analysis it is
clear that, due to the accepted ordering of $V_j$'s, the double
eigenvalues should be identified with the values of the middle
characteristic velocity $V_2$  in the harmonic and the soliton
limits. So we define the boundaries $x^{\pm}(t)$ of the KdV
modulation solution by the characteristic equations (cf.
(\ref{b0}), (\ref{bs}))
\begin{equation}\label{bol}
dx^-/dt=V_2(\bar u, k, 0)|_{x=x^-} \, , \qquad dx^+/dt=V_2( \bar
u, 0, a)|_{x=x^+}\, .
\end{equation}
For the decay of an initial discontinuity Eqs.~(\ref{bol}), in
view of (\ref{bc3}), assume the form
\begin{equation}\label{bs1}
s^-=V_2(u^-, k^-,0)\, , \qquad s^+=V_2(u^+, 0, a^+)
\end{equation}
 where the values $a^+$ and $k^-$ are to be found from
the solution of the GP problem.

In the subsequent sections we will find simple effective
expressions for double characteristics of the Whitham system using
physical variables $\bar u, k, a$ and bypassing the full
eigenvalue analysis. Moreover, we will show that the edge
parameters $a^+$, $k^- $ and hence $s^+$, $s^-$ can be found from
a certain set of conditions {\it not involving the global
integration of the Whitham equations}. As a result, we will
present a way to ``fit'' the dispersive shock into the solution of
the dispersionless limit equations without the detailed analysis
of its internal oscillatory structure in the same manner as the
traditional shock is embedded in the solution of the inviscid
equations of ideal gas. As will follow from the construction, the
availability of these conditions does not depend on the existence
of the Riemann invariants for the Whitham system.

\subsection{Trailing edge}
We start with the determination of the trailing edge speed $s^-$
for the problem of the decay of an initial discontinuity
(\ref{disc}) in the KdV equation using the reformulation of the GP
problem in physical variables made in the previous subsection. The
trailing edge is defined for the solution of the GP problem by the
characteristic where $a=0$. Our analysis below will be based on
the special features of the Cauchy data prescription on
characteristics (see e.g. \cite{ry78}).

It follows from the characteristic form (\ref{char}) that the
differentials of the modulation variables along the $i$-th
characteristic are not independent but related by means of the
 expression
\begin{equation}\label{char1}
b_1^{(i)}(\bar u, k, a)d_i \bar u+ b_2^{(i)}(\bar u, k, a)d_i k +
b_3^{(i)}(\bar u, k, a)d_i a=0 \, .
\end{equation}
Let $a=0$ on this characteristic (which necessarily implies that
it is a double characteristic -- see explanation in the previous
section). Since $a=0$ must be an exact solution of the modulation
system, the ordinary differential equation
\begin{equation}\label{pfaff}
b_1^{(i)}(\bar u, k,0)d_i \bar u + b_2^{(i)}(\bar u, k, 0)d_i k=0
\end{equation}
necessarily represents a  characteristic equation for the
reduction as $a=0$ of the full modulation system. This equation
can always be integrated using an integrating factor to give the
relationship between admissible values of $\bar u$ and $k$ {\it on
the characteristic}, i.e:
\begin{equation}\label{int0}
\Phi_i(\bar u, k)=C_0\, ,
\end{equation}
$C_0$ being  constant of integration. The point is that this
relationship is {\it local}, so it does not depend on the specific
solution under study but is determined (up to a constant) only by
the coefficients of the Whitham system evaluated for $a=0$.
Therefore, the integral (\ref{int0}) can be derived, bypassing the
technically involved route via full characteristic form
(\ref{char1}),  by a direct substitution of the ansatz $k=k(\bar
u)$  into the reduction of the Whitham system for $a = 0$. We also
remark that the value $C_0$ represents, in fact, a Riemann
invariant for this $2\times 2$ reduction. Of course, the existence
of this Riemann invariant does not depend on the existence of the
Riemann invariants for the full Whitham system (\ref{avkdv}).

It follows from the  relationship (\ref{311}a) and Eqs.~(\ref{L}),
(\ref{mean}) considered for $u_2 = u_3$, that
\begin{equation}\label{33}
\lim \limits_{a \to 0} \overline{u^2}=\bar u^2 \, , \quad \lim
\limits_{a \to 0} \omega= \lim \limits_{u_2 \to u_3}ck= k\bar u -
k^3 \equiv \omega_0(\bar u, k ) \, .
\end{equation}
Obviously, the second expression in Eq.~(\ref{33}) represents the
KdV linear dispersion relation, where the linearization is made
about the slowly varying mean background $\bar u (x,t)$. Now we
can immediately obtain the reduction of the Whitham equations
(\ref{wh2})
\begin{equation}\label{wh0n}
 a=0 \, , \quad  \bar u_t+ \bar u  \bar u_x =0 \, , \quad k_t +
(\omega_0 (\bar u, k))_x=0\, .
\end{equation}
The equations (\ref{wh0n}) form a {\it closed} system.  We
emphasize that the system (\ref{wh0n}) is not an asymptotic
system, it is an {\it exact reduction} of the full modulation
system. Of course, the reduction (\ref{wh0n}) is equivalent to the
limiting diagonal system (\ref{wh0}) (which is readily verified by
a direct calculation) but here it was derived directly from the
original modulation equations (\ref{wh2}) without using the
Riemann invariant form. Actually, the limit (\ref{wh0n}) has a
clear meaning of the formal modulation system for a zero-amplitude
wave packet propagating on a slowly varying  background flow $\bar
u$ and can be postulated on the physical level of reasoning
directly, even without turning to the Whitham equations
(\ref{wh2}) and the formal derivation of the linear dispersion
relation (\ref{33}) for modulated waves via nonlinear travelling
wave solution. Now, looking for the integral $k(\bar u)$ of
Eqs.~(\ref{wh0n}) we arrive at the ordinary differential equation
(which is an equivalent of Eq.~(\ref{pfaff})),
\begin{equation}\label{trkdv}
 \qquad \frac{dk}{d\bar u}=\frac{\partial \omega_0 / \partial
\bar u}{\bar u - \partial \omega_0 / \partial k} \, ,
\end{equation}
which, upon substituting the linear dispersion relation from
(\ref{33}) is readily integrated to give
\begin{equation}\label{ku}
    k=\sqrt{\frac{2}{3}(\bar u + C_1)}\, .
\end{equation}
Thus Eq.~(\ref{ku}) is a relationship between values of $k$ and
$\bar u$ {\it on the characteristic} of the Whitham system on
which $a=0$ (there is no relation to any particular global
solution yet).

Now we apply the relationship (\ref{ku}) to the GP problem
(\ref{bc3}), where the characteristic where $a=0$ is associated
with the trailing edge $x=s^-t$. First, we find the constant $C_1$
from the second boundary condition (\ref{bc3}) which, in the space
of the field variables prescribes $k=0$ when $\bar u=u^+$ (this
condition does not contain the amplitude $a$ so it must hold
everywhere in the plane $k=0$ of the three-dimensional space with
the coordinates $\{\bar u, k, a\}$, including the line  $(\bar u,
0,0)$ as well). Thus $C_1=-u^+$. Then, putting $\bar u=u^-$ in
(\ref{ku}) we obtain the value of the wavenumber at the trailing
edge of the dispersive shock $x=s^-t$,
\begin{equation}\label{k02}
k^-=\sqrt{2\Delta/3} \, ,
\end{equation}
where $\Delta=u^--u^+$ is the jump across the dispersive shock.
This agrees with the value of $k^-$ given by Eq.~(\ref{k0})
obtained as a consequence of the full modulation solution
(\ref{vs}).

Now we find the self-similar co-ordinate (the speed) of the
trailing edge $s^-$ which, according to (\ref{bol}), is calculated
as the double characteristic velocity of the Whitham system for
the  solution of the GP problem.  It readily follows from the
limiting system (\ref{wh0n}), that its characteristic velocities
are
\begin{equation}\label{vn}
 \bar u \, , \qquad \frac{\partial
\omega_0}{\partial k}(\bar u, k)=\bar u-3k^2 \, .
\end{equation}
The latter, of course, is the group velocity  of the linear wave
packet propagating on the varying mean flow background $\bar
u(x,t)$, which is perfectly reasonable from physical point of
view. Now we need to identify one of the velocities (\ref{vn})
with the double eigenvalue $V_2(\bar u, k, 0)$ in (\ref{bs1}a) to
evaluate the trailing edge speed. It is clear from physical
reasoning that this must be the linear group velocity. Later we
will present additional formal conditions eliminating any possible
ambiguity in the identification of the double eigenvalue. Of
course, for the KdV case it is not difficult to show using
explicit formulas (\ref{mean}), (\ref{L}) that expressions
(\ref{vn}) are equivalent to those obtained from the Riemann form
of the Whitham equations in the harmonic limit (see
Eq.~(\ref{hv})) and our identification $V_2(\bar u, k, 0)=\partial
\omega_0/\partial k$ is indeed correct. Thus, we get from
Eq.~(\ref{bs1}a) for the trailing edge
\begin{equation}\label{tre}
s^-=\frac{\partial \omega_0}{\partial k}(u^-,k^-) \, .
\end{equation}
In this form the equation of the trailing edge can be interpreted
as a kinematic boundary condition for linear modulated waves and
can be postulated as a part of the problem formulation.

Setting (\ref{vn}), (\ref{k02}) into (\ref{tre}) we obtain for the
trailing edge
\begin{equation}\label{s-}
s^-= u^--3(k^-)^2=u^+-\Delta\, .
\end{equation}
This value, again, agrees with the full solution of the GP problem
(see Eq.~(\ref{spm0})).

\subsection{Leading edge}
In principle, one could proceed with the leading edge in the same
fashion as we did with the trailing edge i.e. by indirect
integration of the full characteristic $1$-form along a
characteristic where $k=0$, where this form degenerates into
\begin{equation}\label{pfaff1}
b_1^{(i)}(\bar u, 0,a)d_i \bar u + b_2^{(i)}(\bar u, 0, a)d_i
a=0\, .
\end{equation}
That would imply finding the integral $a(\bar u)$ for the exact
zero-wavenumber reduction of the modulation system. Such a
reduction, however, must include full modulation equation for the
soliton amplitude, which can be obtained without too much trouble
for the KdV equation but is not readily available for more
complicated systems, especially if the solitary wave solution can
not be found explicitly. One should remark that the ``obvious''
universal soliton amplitude equation $a_t+c_s(a)a_x=0$ (see
\cite{Wh74}, Sec.~16.6) is correct only if the background flow
$\bar u$ is constant. The full amplitude equation for solitary
waves (see \cite{grim79}, \cite{gke90} for instance) takes into
account variations in all modulation parameters and inevitably
contains the term proportional to $\bar u_x$ which is vital for
deriving the characteristic equation (\ref{pfaff1}). As a result,
unlike the wave number conservation law, the {\it full} soliton
amplitude equation does not have an explicit universal form and a
straightforward application of the method proposed in Section 3.2
would be essentially equivalent to the consideration of the
general characteristic equation
 (\ref{char1}) in the soliton limit. For ``real'' non-integrable systems
 that
 could  be  a very difficult technical task. This complication, however, can be bypassed by introducing
a conjugate wavenumber defined for the KdV as
\begin{equation}\label{ck}
    \tilde k(u_1,u_2,u_3)= \pi
\left(\int\limits^{u_2}_{u_1}\frac{du}{\sqrt{G(u)}}\right)^{-1}=\frac{\pi}{2\sqrt{3}}
\frac{(u_3-u_1)^{1/2}} { K(m')}\, , \quad m'=1-m \,
\end{equation}
-- instead of the amplitude $a$ and the ratio $\Lambda=k/\tilde k$
instead of the original wavenumber $k$ (\ref{L}). The new set of
modulation variables we are going to use is ($\bar u, \Lambda,
\tilde k$). It is readily seen that the  soliton limit $k = 0$ ($m
= 1$) corresponds to $\Lambda = 0$, $\tilde k =
\sqrt{a_s/3}=\mathcal{O}(1)$ in new variables. On the other hand,
in the harmonic limit $a=0$ ($m=0$) one has $\tilde k =0$. So
$\tilde k$ indeed plays the role analogous to the amplitude. For
 further convenience, we rewrite the matching conditions
(\ref{bc3}) using new variables
\begin{equation}
\begin{array}{l}
x=s^-t:\qquad \tilde k=0\, , \ \ \bar u = u^- \, , \\
x=s^+t:\qquad \Lambda=0\, , \ \ \bar u = u^+ \, .
\end{array}
\label{bc4}
\end{equation}
Now, as was shown in Section 3.1, the two first averaged
conservation laws (\ref{wh2}) in the soliton limit reduce to the
Hopf equation:
\begin{equation}\label{wh10}
 \Lambda=0 : \qquad  \bar u_t + \bar u \bar u_x=0 \, .
\end{equation}
To obtain the equation for $\tilde k$ in the soliton limit we
 set the new variables in the wave number conservation law in
 (\ref{wh2}) to obtain:
\begin{equation}\label{55}
\tilde k  \Lambda_t + \tilde \omega \Lambda_x+ \Lambda ( \tilde
k_t +  \tilde \omega_x )=0 \, ,
\end{equation}
where $\tilde \omega=\tilde \omega(\bar u, \Lambda, \tilde k)  =
\tilde k c$ is the conjugate frequency.

Now, using the arguments identical to those in the previous
subsection we infer that if $\Lambda =0$ on some characteristic
then the values of $\tilde k$ and $\bar u$ on this characteristic
must be connected by a local relationship $ \tilde k (\bar u)$
which must not depend on the particular solution. We consider
equation (\ref{55}) in the small vicinity of the mentioned
characteristic where $\Lambda \ll 1$. Then, using $\tilde k =
\tilde k(\bar u)$ and Eq.~(\ref{wh10}) for the leading order, we
obtain an asymptotic equation
\begin{equation}\label{56}
 \frac{\partial \Lambda}{\partial t} + \frac{\tilde
\omega_s}{\tilde k} \frac{\partial \Lambda}{\partial
x}+\frac{\Lambda}{\tilde k}\frac{\partial \bar u}{\partial x}
\left\{ \frac{d \tilde k}{d \bar u}\left(\frac{\partial \tilde
\omega_s}{\partial \tilde k}-\bar u \right)+\frac{\partial \tilde
\omega_s}{\partial \bar u} \right\}= \mathcal{O}\left(\Lambda
\frac{\partial \Lambda}{\partial x}\right) \, ,
\end{equation}
where the reduction $\tilde \omega_s (\bar u, \tilde k)=\tilde
\omega(\bar u, 0, \tilde k)$ can be called a {\it soliton
dispersion relation}. Since the sought characteristic integral
$\tilde k(\bar u)$ must not depend on the way $\Lambda$ tends to
zero, the expression in the brackets (which does not depend on
$\Lambda$) must be identically zero and Eq.~(\ref{56}) splits into
\begin{equation}\label{lkdv0}
 \frac{d \tilde k}{d\bar u}=\frac{\partial \tilde \omega_s /
\partial \bar u}{\bar u - \partial \tilde \omega_s / \partial \tilde k}
\end{equation}
and
\begin{equation}\label{57}
\qquad \frac{\partial \Lambda}{\partial t} + \frac{\tilde
\omega_s}{\tilde k}\frac{\partial \Lambda}{\partial x}=
\mathcal{O}\left(\Lambda \frac{\partial \Lambda}{\partial
x}\right) \, .
\end{equation}
Eqs.~(\ref{lkdv0}), (\ref{57})  have been obtained in \cite{ekt03}
using somewhat more particular arguments. One can not help
noticing that the equations (\ref{trkdv}) and (\ref{lkdv0}) for
the characteristic integrals $k(\bar u)$ (linear characteristic)
and $\tilde k(\bar u)$ (soliton characteristic) are identical in
terms of the corresponding dispersion relations $\omega_0(\bar
u,k)$ and $\tilde \omega_s(\bar u, \tilde k)$. The latter,
however, is yet to be found.

It follows from  (\ref{57}), (\ref{wh10}) that in the soliton
limit the characteristic velocities of the Whitham system are
\begin{equation}\label{vs}
 \frac{\tilde \omega_s}{\tilde k}\, , \qquad \bar u \, .
\end{equation}

One can see from  the definition of $\tilde \omega_s$ that the
characteristic velocity  $\tilde \omega_s/\tilde k $ coincides
with  the soliton speed $c_s(\bar u, \tilde k)=c(u_1,u_1,u_3)$,
which is, of course,  expected. We identify $c_s$ with the double
characteristic velocity $V_2=V_3$ of the full Whitham system in
the soliton limit, then the classical speed $\bar u$ would
correspond to $V_1$. The conditions eliminating possible ambiguity
in the characteristic velocity identification for the reduced
Whitham system will be presented later. Now, similarly to the
trailing edge case, the definition of the leading edge in
Eq.~(\ref{bs1}) can be represented as a kinematic boundary
condition for the lead soliton (cf. (\ref{tre}) )
\begin{equation}\label{le}
s^+=c_s(u^+, \tilde k^+) \, ,
\end{equation}
where $\tilde k^+=\tilde k(u^+)$. One can see that in this  form
the definition of the leading edge can be adopted as a part of the
problem formulation.

 At this point, one may take advantage of the
explicit expressions (\ref{mean}), (\ref{L}) for $\bar
u(u_1,u_2,u_3)$, $\tilde k (u_1,u_2,u_3)$ and $c(u_1,u_2,u_3)$ to
find $\tilde \omega_s(\bar u, \tilde k)$ and  solve Eq.
(\ref{lkdv0}) using the matching conditions (\ref{bc4}). Such an
attractive route, however, might not be readily (if at all)
available for actual non-integrable systems, where the physical
modulation parameters often can not be expressed in a simple way
in terms of the roots of the potential curve. So we proceed with
some apparently "roundabout" way, which would be more universally
applicable to other systems.

We observe that $\tilde k$ and  $\tilde \omega=\tilde k c$ defined
by (\ref{ck}) and (\ref{L}a) as functions of the roots
$u_1,u_2,u_3$ can be viewed as the wavenumber and the frequency in
the {\it conjugate travelling wave} associated with the same set
of roots $u_j$ as in Eq.~(\ref{trcn}) but inverted potential curve
(now oscillations occur between the roots $u_1$ and $u_2$):
\begin{equation}\label{trcnc}
(\tilde u_{\tilde \theta})^2= G(\tilde u)\, , \qquad \tilde u
(\tilde \theta + 2\pi/\tilde k) = \tilde u (\tilde \theta) \, .
\end{equation}
where $\tilde u$ is new field variable and $\tilde \theta = \tilde
x - c \tilde t$ is a new travelling phase associated with ``old"
phase velocity $c=(u_1+u_2+u_3)/3$. Eq.~(\ref{trcnc}) can be
obtained from (\ref{trcn})  by the change of variables $u \mapsto
\tilde u$, $x \mapsto i\tilde x$, $t \mapsto i \tilde t$, which
corresponds to a mere change of the dispersion sign in the KdV
equation (\ref{kdv}),
\begin{equation}\label{ckdv}
\tilde u_{\tilde t}+ \tilde u \tilde u_{\tilde x}-\tilde u_{\tilde
x \tilde x \tilde x}=0 \, .
\end{equation}
As a matter of fact, in our particular KdV case the functions
$u(\theta)$ and $i\tilde u(-i\tilde \theta)$ defined for the same
set of roots $u_j$ represent the same analytic (elliptic) function
in the complex $x$-plane with the periods $2\pi/k$ and $2\pi i
/\tilde k$ along the real and the imaginary axes. Generally, of
course, this is not the case, so we are not going to take
advantage of the analytic properties of $u(\theta)$ here. Instead,
we observe that the soliton limit for the original travelling wave
equation (\ref{trcn}) ($u_2 \to u_1$) corresponds to the harmonic
limit for the conjugate equation (\ref{trcnc}). Therefore, the
relation between $\tilde \omega_s $ and $\tilde k$ can be obtained
as a {\it linear dispersion relation} for the conjugate KdV
equation (\ref{ckdv}), i.e.
\begin{equation}\label{cdr0}
\tilde \omega_s=\tilde k<\tilde u>+ \tilde k^3 \, ,
\end{equation}
where the brackets $<...>$ denote the  averaging over the
conjugate family (\ref{trcnc}) (cf. (\ref {av}))
\begin{equation} \label{avc}
  <F>(u_1,u_2,u_3)=  \frac{\tilde k}{\pi}\int
\limits^{u_2}_{u_1}\frac{F(\tilde u)}{\sqrt{G(\tilde u )}}d \tilde
u \, .
\end{equation}
It is not difficult to show by a direct calculation that
$<F>(u_1,u_1,u_3)=F(u_1)$ . Then, it follows from Eq.~(\ref{31})
that for $u_2=u_1$ ($k=0$) we have
\begin{equation}\label{prop}
<F(u)>|_{k=0}=\overline{F(u)}|_{k=0}
\end{equation}
 and, in particular, $<u>=\bar u$. Again, similarly to Eqs.~(\ref{31}), (\ref{311})
 this property is due to the quadratic behaviour of the
 potential curve $G(u)$ (\ref{Q})in the vicinity of the double root corresponding to the soliton limit and
is not confined to the KdV example alone.

It follows from Eq.~ (\ref{prop}) that  $\tilde \omega_s(<u>,
\tilde k)=\tilde \omega_s(\bar u, \tilde k)$ and therefore, the
soliton dispersion relation can be obtained from the original
linear dispersion relation $\omega_0(\bar u, k)$ by the formal
change
\begin{equation}\label{mapsto}
    k \mapsto i\tilde k \, , \qquad \omega_0 \mapsto i \tilde
    \omega_0
\end{equation}
In other words, the soliton dispersion relation is found as
\begin{equation}\label{sdr}
\tilde \omega_s(\bar u, \tilde k)=-i\omega_0(\bar u, i\tilde k
)=\tilde k \bar u  +\tilde k^3\, .
\end{equation}
Of course, in view of  $\tilde k (u_1,u_1, u_3)=\sqrt{a_s/3}$ and
$\bar u (u_1,u_1, u_3)=u_1$, Eq.~(\ref{sdr})  is equivalent to the
classic relationship (\ref{sol00}) between the speed  and the
amplitude of the KdV soliton but here it was obtained in terms of
the linear dispersion relation so the outlined procedure can be
easily generalised to more complicated systems, where the explicit
analysis of the travelling wave solution is not as readily
available as in the KdV case.

 The possibility of expressing the soliton speed in
terms of the linear dispersion relation might look somewhat
surprising but can be  explained by the following simple argument
(see \cite{kam03}): the value of $u$ in the soliton tail, which
moves with the same speed as the rest of the soliton, is very
small, so one can, in principle, infer this speed from the linear
theory.

{\bf Remark. } We emphasize that  the obtained relationships
between the original and the conjugate averaged variables
essentially represent algebraic identities between integrals of
the form (\ref{av}) and (\ref{avc}) associated with a given
Riemann surface $\mu^2=G(u)$ and do not imply any connection
between the spatio-temporal modulation dynamics for the original
and the conjugate equations (\ref{kdv}) and (\ref{ckdv}).

 Now setting Eq.~(\ref{sdr}) into
Eq.~(\ref{lkdv0}) we obtain after elementary integration
\begin{equation}\label{kul}
    \tilde k=\sqrt{\frac{2}{3}(C_2-\bar u )}\, ,
\end{equation}
where $C_1$ is constant of integration. We emphasize that
Eq.~(\ref{kul}) (as well as Eq.~(\ref{ku}) in the linear case ) is
a general relationship between values of $\tilde k$ and $\bar u$
on the characteristic of the Whitham system where $\Lambda=0$ and
is not tied to any particular global solution.

Now we apply the relationship (\ref{kul}) to the GP problem
(\ref{bc4}), where the characteristic on which $\Lambda=0$ is
associated with the leading edge $x=s^+t$. First, we find the
constant $C_2$ from the first boundary condition (\ref{bc4}) which
prescribes $\tilde k=0$ when $\bar u=u^-$ in the three-dimensional
space with the coordinates $\{\bar u, \Lambda, \tilde k\}$. This
condition does not contain
 $\Lambda$ so it must hold on the line $\{\bar u, 0,
0 \}$  as well and, therefore, can be applied to (\ref{kul}). Thus
$C_2=u^-$. Then, putting $\bar u=u^+$ in (\ref{kul}) we obtain the
value of the conjugate wavenumber at the trailing edge of the
dispersive shock $s=s^+$:
\begin{equation}\label{ks2}
\tilde k^+=\sqrt{2\Delta/3} \, .
\end{equation}
We observe that, for the KdV equation  $\tilde k^+=k^-$ (see
(\ref{k02})). Of course, this is not a general relationship
although one can expect significant symmetry between the
expressions for $k^-$ and $\tilde k^+$ in other
dispersion-hydrodynamic systems.

The speed of the leading edge (\ref{le}) is now calculated with
the aid of the conjugate dispersion relation (\ref{sdr}) as
\begin{equation}\label{}
s^+=\frac{\tilde \omega_s (u^+ , \tilde k^+)}{\tilde k^+}=u^+ +
\frac{2}{3}\Delta \, ,
\end{equation}
which agrees with the  global  solution of the full Whitham system
(see Eq.~(\ref{spm0}) ). The KdV  soliton amplitude is connected
with its velocity by the relation (\ref{sdr0}), which in our case
assumes the form $s^+=u^+ + a^+/3$. Hence the lead soliton
amplitude is $a^+=3(s^+-u^+)=2\Delta$, which,  again, agrees with
the full modulation solution.
\subsection{``Entropy'' conditions.}
One can see that the  valid solution to the GP problem with
$u^->u^+$ must satisfy the inequalities
\begin{equation}\label{in0}
 s^-<u^- \, , \qquad s^+>u^+\, ,\qquad  s^+>s^- \, ,
\end{equation}
These inequalities ensure that the ``external", Hopf
characteristics $x=u^-t$ and $x=u^+t$ starting from the $x$-axis
on either side of the dispersive shock  region intersect  its
edges when continued in the direction of increasing $t$ thus
transferring the initial data {\it into} the dispersive shock zone
$s^-t<x<s^+t$ (see Fig. 3).  The inequalities (\ref{in0})
represent a dispersive- hydrodynamic  analog of the classical gas-
dynamic {\it entropy conditions} \cite{lax73}. As a matter of
fact, these inequalities are redundant when the explicit solution
of the GP problem is available (one can see that conditions
(\ref{in0}) are indeed satisfied by the boundaries of the solution
(\ref{simkdv})). However, in the absence of the full modulation
solution, and without physical assumptions about the spatial
structure of the dispersive shock there is an ambiguity in the
determination of the edges using the construction proposed in
Sections 3.2, 3.3. This ambiguity has been pointed out when we
made an identification of the leading and trailing edges with the
soliton and the harmonic wave  in the natural boundary conditions
(\ref{bc3}) and when we identified linear group velocity and the
soliton velocity with the double eigenvalues of the full Whitham
system in respective limits.
\begin{figure}[ht]
\centerline{\includegraphics[width=8cm,height=
6cm,clip]{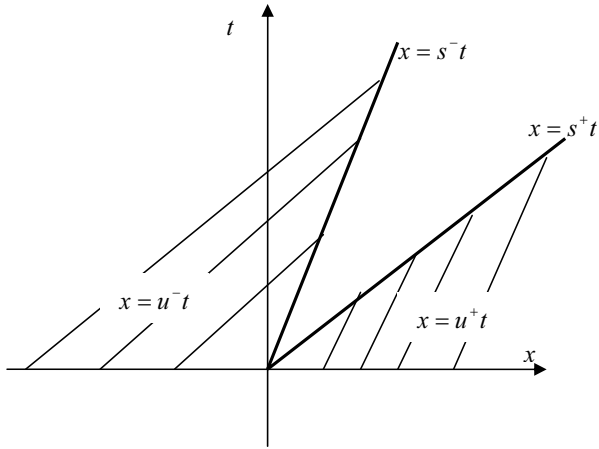}} \vspace{0.3 true cm} \caption{Classical
(Hopf) characteristics transfer initial data into the Whitham
zone} \label{fig3}
\end{figure}
One may speculate about the possible analog of the entropy for
dispersive hydrodynamics. An attractive candidate is given by the
integral
\begin{equation}\label{ent}
    S\propto \int \limits_{-\infty}^xkdx' \ge 0\, .
\end{equation}
For the KdV dispersive shock  the {\it positive} function $k(x,t)$
is supported on the interval $s^-t<x<s^+t$ and one can see that
the integral (\ref{ent}) increases when crossing the dispersive
shock.

\subsection{Geometric interpretation }

We now place  the obtained relationships in a  more general
context. The quasilinear hyperbolic system of the third order
(\ref{ql}) admits the centred expansion fan solution
\begin{equation}\label{int}
F_1(\bar u, k, a)= I_1\, , \ F_2(\bar u, k,a)= I_2\, , \ V_k(\bar
u, k, a)= x/t\, ,
\end{equation}
$I_1$, $I_2$ being constants and  $V_k$ is one of the
characteristic velocities  so that the solution satisfied the GP
matching conditions (\ref{bc3}) (we know that for the KdV equation
this is $V_2$ -- see Eq.~(\ref{simkdv})). The two first
expressions in (\ref{int}) define, for given $I_1, I_2$, two
integral surfaces in the space of the field variables $\bar u, k,
a$ with $k>0$, $a>0$ (physical restrictions). Their intersection
yields the solution curve $\{\Gamma(s;I_1,I_2): \ \ \bar u= \bar
u(s)\, , k=k(s)\, , a=a(s)\}$, where the parameter $s=x/t$ changes
on the interval $[s^- ,s^+]$ so that $a(s^-)=0$ and $k(s^+)=0$.

We now briefly outline  how one proceeds with this geometric
construction. First we observe that, the endpoints of the solution
curve $\Gamma (s)$ lie on the lines of intersection of two
integral surfaces parametrised by $I_1$ and $I_2$ with the
coordinate planes $a=0$ and $k=0$. It follows from (\ref{int})
that there are four  these lines: $F_{1,2}(\bar u, k, 0)= I_{1,2}$
and $F_{1,2}(\bar u, 0, a)= I_{1,2}$ . Since the Whitham system
admits exact $2 \times 2$ hyperbolic reductions for $a=0$ and
$k=0$, the equations of the intersection lines can be obtained
directly, as the characteristic integrals (Riemann invariants) of
the reduced systems, i.e as solutions in the form $k(\bar u)$ and
$a(\bar u)$ respectively. The constants $I_{1,2}$ are then
expressed in terms of the initial parameters $u^-$, $u^+$ by
applying the boundary conditions (\ref{bc3}) to the characteristic
integrals as it was done in Sections 3.2, 3.3. Then, using the
kinematic conditions (\ref{tre}), (\ref{le}), one evaluates the
speeds of the trailing and the leading edges and only after that,
the valid pair of the endpoints is selected by the ``entropy''
inequalities (\ref{in0}).

Essentially, this general construction has been realised in
Sections 3.2, 3.3 by using the arguments from the characteristics
theory and introducing a different (conjugate) basis of the field
variables when considering the leading edge.
\section{Dispersive shocks in simple-wave led dispersive equations}
It is clear that the presented construction is not restricted by
the KdV equation alone, and  can be naturally generalised to the
"KdV-like" nonlinear dispersive systems possessing the basic
properties necessary for  the Whitham averaging.
 We represent the governing equation in the form
\begin{equation}\label{single}
u_t+V(u)u_x+K_3[u]=0\, ,
\end{equation}
where $V(u)$ is a real function and  $K_3$ is a real differential
operator of the third order  with respect to  spatial or mixed
derivatives such that the equation (\ref{single}) has real linear
dispersion relation $\omega=\omega_0(k)$. We assume the following
general properties for the equation (\ref{single}):

 (i) it admits the hyperbolic
classical (dispersionless) limit obtained formally by introducing
the stretched independent variables $x' =\epsilon x$, $t'=
\epsilon t$ and tending $\epsilon \to 0$ while assuming finiteness
of the derivatives with respect to $x'$ and $t'$,
\begin{equation}\label{Rw}
u_{t'} + V(u)u_{x'}=0\, .
\end{equation}
In terms of the linear dispersion relation this property implies
$\omega_0 \sim k$ for $k \ll 1$ and is associated with weakly
dispersive waves.

 (ii) it possesses at least  two conservation laws;

(iii) it supports periodic travelling waves parametrised by three
independent integrals of motion such that the travelling wave
solution allows for a harmonic (zero-amplitude) and a solitary
wave (zero wavenumber) limits. We will assume the following
KdV-like behaviour for the ``potential'' function $G(u)$ in the
ordinary differential equation $(u_\theta)^2= -G(u)$ specifying
the travelling wave solution (cf. Eq.~(\ref{Q})): a) The function
$G(u)$ has at least three real zeros $u_3 \ge u_2 \ge u_1$ such
that the oscillations occur between $u_2$ and $u_3$ (the latter is
assumed just for sake of definiteness); b) In both nearly linear
$(u_3-u_2)/(u_3-u_1)\ll 1$ and nearly soliton
$(u_2-u_1)/(u_3-u_1)\ll 1$ configurations the generic (quadratic)
asymptotic behaviour is assumed:
\begin{equation}\label{quadr}
 G(u) = (u-u_1)(u-u_2)(u_3-u)G^*(u)\, ,
  \end{equation}
 where
 \begin{equation}\label{Gu}
 G^*(u)=\mathcal{O}(1) \quad \hbox{for} \quad \left|\frac{u-u_2}{u_3-u_1}\right| \ll 1
\end{equation}
so that the limiting transitions (\ref{31}), (\ref{prop}) for the
mean values can be easily shown to take place;

(iv) the corresponding Whitham system of the third order (two
averaged ``hydrodynamic'' conservation equations plus the wave
number conservation law) is hyperbolic for the solutions under
study.

The equations of the form (\ref{single}) possessing the mentioned
properties (i) - (iv) may be characterised as {\it simple-wave led
weakly dispersive}. We note that all assumptions except (iv) can
usually be explicitly verified for a specific system. We partially
address the hyperbolicity issue (iv) in the end of this section.

 We consider initial data in the form of an arbitrary step
\begin{equation}\label{discsingle}
u_0(x)=u^- \quad  \hbox{for} \ x<0\, ; \qquad u^+  \quad
\hbox{for} \ x>0\, , \ .
\end{equation}
and assume that the dispersive shock transition for $t >0$ can be
modelled by the centred expansion fan solution to the Whitham
equations. Now one can see that all  arguments used in the
derivation of the dispersive shock conditions for the KdV equation
in a ``non-integrable'' reformulation hold for the case being
considered: one just needs to replace the Hopf ``dispersionless''
speed with $V(u)$ and the KdV linear dispersion relation with the
dispersion relation corresponding to (\ref{single}). So we present
only the final relationships without making any assumptions about
mutual position of the soliton and harmonic edges.

First we introduce the linear dispersion relation for
Eq.~(\ref{single}) by considering an infinitesimal perturbation of
a mean level $\bar u$
\begin{equation} \label{lin}
u\approx \bar u +u_1 e^{i(kx-\omega t)}\, ,  \quad  u_1 \ll 1 \, ,
\end{equation}
which yields   $\omega = \omega_0(\bar u, k)$ to  leading order.
Let the dispersive shock transition be confined to an interval
$s^-t \le x \le s^+ t$. We introduce two sets of parameters
$\{k^-, s^-;\tilde k^+, s^+\}_1$ and $\{\tilde k^-, s^-;k^+,
s^+\}_2$ . The set $\{k^-, s^-;\tilde k^+, s^+\}_1$ is associated
with the negative dispersion wave pattern when the soliton appears
at the leading edge of the dispersive shock. We define this set in
the following way. First we find two functions $k(\bar u)$ and
$\tilde k(\bar u)$ from the ordinary differential equations
\begin{equation}\label{ode1}
\frac{dk}{d\bar u}=\frac{\partial \omega_0 / \partial \bar
u}{V(\bar u) - \partial \omega_0 / \partial k} \, , \qquad
k(u^+)=0\, ,
\end{equation}
\begin{equation}\label{ode2}
 \frac{d \tilde k}{d\bar u}=\frac{\partial
\tilde \omega_s /
\partial \bar u}{V(\bar u) - \partial \tilde \omega_s / \partial \tilde k} \,
,\qquad \tilde k(u^-)=0 \, .
\end{equation}
Then the values of the wavenumber at the trailing edge $k^-$ and
the lead soliton conjugate wavenumber $\tilde k^+$ are  found as
  $k^-=k(u^-)$, $\tilde k^+ = \tilde k(u^+)$. By definition, they
  must be real. The speeds of the dispersive shock edges are found
  from the expressions
\begin{equation}\label{spm}
s^-=\frac{\partial \omega_0 }{\partial k}(u^-, k^-)\, , \qquad
s^+=\frac{\tilde \omega_s(u^+, \tilde k^+ )}{\tilde k^+} \, ,
\end{equation}
 where $\tilde \omega_s(\bar u, \tilde k )=-i \omega_0(\bar u, i \tilde
 k)$.
The second set of parameters $\{\tilde k^-, s^-;k^+, s^+\}_2$
associated with the reversed, positive dispersion pattern, is
obtained from the same system (\ref{ode1}) -- (\ref{spm}) in which
one replaces ``$-$" with ``$+$" which is equivalent to a
respective replacing of the parameter subscripts directly in the
set $\{\cdot\}_1$. The valid set is selected by the reality
condition for the wavenumber and by the ``entropy" conditions,
\begin{equation}\label{ent}
s^-<V(u^-)\, ,  \qquad s^+>V(u^+) \, , \quad s^+ > s^- \, .
\end{equation}
In principle, the solution can switch between the sets
$\{\cdot\}_1$ and $\{\cdot\}_2$ depending on the initial
conditions. ``Switching'' of the edge parameter sets implies the
reversion of the spatial structure of the dispersive shock.
Another possibility is that for some domain of the initial data
$(u^+, u^-)$ both sets $\{\cdot\}_1$ and $\{\cdot\}_2$ fail to
satisfy either reality or ``entropy'' condition. This would imply
that  a genuine global solution to the dispersionless limit
equations is available for such initial data and no breaking
occurs. This solution, of course, is a classical rarefaction wave.

We also briefly address the issue of the modulational stability.
Our main mathematical assumption in this paper is that of the
hyperbolicity of the Whitham system for the solutions under study.
This ensures {\it global} modulational stability of the dispersive
shock. For non-integrable dynamics, establishing the region of
hyperbolicity for the modulation system is a separate, often
technically involved, problem. However, in a more restricted
context of the dispersive shock description, some effective
necessary conditions of global modulational stability can be
formulated in the following way.

 We evaluate the the frequency and conjugate frequency
at the respective boundaries of the dispersive shock transition
using the solutions of the ordinary differential equations
(\ref{ode1}), (\ref{ode2}). We denote these frequencies as
$\omega_0[u^-, u^+]$ and $\tilde \omega_s[u^-, u^+]$. Then, given
the strict hyperbolicity of the dispersionless limit, the
necessary criterion for global modulational stability of the
dispersive shock is given by the conditions
\begin{equation}\label{stab}
\hbox{Im} \  \omega_0[u^-, u^+]=0 \, , \qquad \hbox{Im} \ \tilde
\omega_s[u^-, u^+]=0 \, ,
\end{equation}
which are equivalent to the natural requirement for the speeds
$s^-$, $s^+$ to be real. The conditions (\ref{stab}) define the
domain $D$ in the  initial data plane $(u^+, u^-)$ which
corresponds to modulationally stable solutions.  It is not clear
if the formulated criterion  is the sufficient condition  so the
obtained region $D[u^+;u^-]$ could be corrected by some additional
conditions.

We note that, for non-integrable equations  some additional
restrictions on admissible values of initial data can occur due to
existence conditions for the single-phase travelling wave
solutions, for instance, due to the presence of the upper bound
for the soliton amplitude.

\vspace{0.5cm} {\bf Example:} {\it Decay of a step problem for the
defocusing mKdV equation}

As the next simplest example  of an effective construction of the
dispersive shock transition in the simple-wave led dispersive
equation  we consider the step resolution problem for the
defocusing mKdV (mKdV(d)) equation
\begin{equation}\label{mkdv}
    u_t-u^2u_x+u_{xxx}=0\, ,
\end{equation}
\begin{equation}\label{discsmkdv}
u(x,0)=u_- \ \hbox{for} \ x<0\, ; \ \  u_+  \ \hbox{for} \ x>0\, .
\end{equation}
Equation (\ref{mkdv}) is an exactly integrable equation and can be
treated by the IST method. It belongs to the defocusing NLS
hierarchy (self-adjoint spectral operator) and  its Whitham system
is known to be hyperbolic \cite{lev99}. Actually, the modulation
system for mKdV(d) equation in Riemann invariants is identical to
that for the KdV equation (this fact had been established  in
\cite{don75} by a direct calculation before the methods of
finite-gap modulation theory became available). That, however,
does not imply that the physical modulation solutions for KdV and
mKdV(d) in terms of initial data $u^+$ and $u^-$ will be the same
or even have same properties. We will demonstrate now how major
qualitative and quantitative characteristics of the modulation
solution can be derived from the dispersive shock conditions
(\ref{ode1}) -- (\ref{spm}) without derivation of the Whitham
system and analysis of its full solution.

We infer from (\ref{mkdv}) that $V(\bar u)=-\bar u^2$,
$\omega_0(\bar u, k)=-k\bar u^2-k^3$. Then using (\ref{ode1}) --
(\ref{spm}) we readily obtain two possible sets for the edge
parameters:
\begin{equation}\label{set1}
k^-_1= \tilde k^+_1=\sqrt{\frac{2}{3}(u_+^2-u_-^2)}\, , \quad
s^-_1=u_-^2-2u_+^2\, , \quad
s^+_1=-\frac{1}{3}u_+^2-\frac{2}{3}u_-^2 \, ,
\end{equation}
\begin{equation}\label{set2}
k^+_2= \tilde k^-_2=\sqrt{\frac{2}{3}(u_-^2-u_+^2)}\, , \quad
s^-_2=-\frac{1}{3}u_-^2-\frac{2}{3}u_+^2 \, , \quad
s^+_2=u_+^2-2u_-^2\, .
\end{equation}
There are three cases to consider.

(i)  Let $u_+^2-u_-^2>0$. Then the second set (\ref{set2}) must be
discarded by failure to satisfy the reality condition for the
wavenumbers. The first set (\ref{set1}) satisfies both the reality
condition and the ``entropy'' conditions. Thus, for
$u_+^2-u_-^2>0$ one gets the dispersive shock with the negative
dispersion wave pattern. The width of the dispersive shock is
$(s^+-s^-)t=\frac{5}{3}(u_+^2-u_-^2)t$. The amplitude of the lead
soliton is found from the relationship between the velocity and
the amplitude for the mKdV(d) solitons moving against the
background $\bar u$, which is obtained by an elementary analysis
of the travelling wave solution in the soliton limit and has the
form : $c_s=-\bar u^2 \pm 2\bar u a_s/3 - a_s^2/6$. Different
signs in this expression correspond to solitons of different
polarity supported by the mKdV(d) equation. Now, after simple
calculation one gets $a^+=2(|u_+| - |u_-|)$. For $u_+u_- \ge 0$
this coincides with the KdV result (\ref{as0}) (to make such a
comparison in the case when $u^+>u^- \ge 0$, the KdV equation
should be taken with the negative sign for the nonlinear term,
which would result in the change $\Delta \to -\Delta$ in
(\ref{as0})). The location of the mKdV(d) dispersive shock given
by (\ref{set1}), however, is different from the respective KdV
case as for the KdV equation with the same initial data one would
have $|s^+-s^-|=\frac{5}{3}(|u_+|-|u_-|)t$ rather than
$\frac{5}{3}(u_+^2-u_-^2)t$.

(ii) $u_+^2=u_-^2$, which implies $u_+=u_-$ (trivial case) or
$u_+=-u_-$. The width of the dispersive shock becomes zero, which
implies that symmetric initial discontinuity with $u_+=-u_-$ does
not break and propagates as a whole with the velocity $-u_+^2$.
This corresponds to the exact solution of the mKdV(d) equation in
the form of a smooth kink (in the Whitham approximation its width
is equal to zero, hence the propagating discontinuity).

(iii) If $u_+^2-u_-^2<0$, the first set of parameters (\ref{set1})
fails to satisfy the reality condition for the wavenumbers while
the second set (\ref{set2}) does not pass the ``entropy'' test.
This implies that there exists a genuine global solution to the
dispersionless limit which does not require a dispersive shock.
This solution is a rarefaction wave $x/t=-u^2$ confined to the
interval $-u^2_-t\leqslant x \leqslant -u^2_+t$.

Of course, these results could be inferred from the known periodic
solution and the full modulation system for the mKdV(d) equation
in Riemann invariants (see \cite{kam2004} for instance) but here
they have been obtained without invoking the integrable structure
mKdV(d) equation, as an illustration of the effectiveness of the
general transition conditions for a dispersive shock.

\section{Bi-directional dispersive hydrodynamics}

\subsection{General setting and formulation of the problem}
We now generalise the obtained results to a physically important
case of  $2\times2$ strictly hyperbolic systems modified
 by weak dispersion. We represent such a
system in a symbolic form
\begin{equation}\label{gen}
{\phi}_t= {\bf K}_{4} (\phi,\phi_x, \phi_{xx}, \phi_{xt}\dots )\,.
\end{equation}
 where $\phi$ is $2$-vector,  ${\bf K}_4$ is  vector differential operator
  of fourth order with respect to spatial/mixed
 derivatives so that the system (\ref{gen})  has real linear dispersion relation
$\omega=\omega_0(k)$ so that $\omega_0 \to k$ as $k \ll 1$ (weak
dispersion). We assume that the system (\ref{gen}) has at least
three independent conservation laws
\begin{equation}\label{cons1}
\frac{\partial P_j}{\partial t} + \frac{\partial Q_j}{\partial
x}=0 \,,  \quad j=1,2,3.
\end{equation}
For convenience of explanation we associate two conserving
densities $P_{1,2}$ with the ``gas'' density $\rho$ and the
momentum $\rho u$ and assume that the dispersionless limit of
(\ref{gen}) has
 the form of the  gas-dynamic Euler equations for the isentropic ideal
 gas
\begin{equation} \label{euler}
 \rho_t + (  \rho u)_x = 0 \, , \qquad (\rho u)_t +  (\rho u^2 + p(\rho))_x = 0 ,
\end{equation}
where $p(\rho)$ is the pressure in the corresponding gas dynamics.
It should be emphasized that generally speaking the assumption
about the gas dynamic `core' of the system (\ref{gen}) is just a
convenient (and in many cases physically relevant) way to convey
our ideas -- actually there is no need to restrict oneself with
this particular form of the dispersionless limit. The only
property of the dispersionless limit system which will be used
below is that it can be represented in the Riemann form (which is
always the case for quasilinear $2\times2$ systems). So actually
one should put quotation marks for the ``density'', ``velocity''
and ``pressure'' in this section.

As earlier, we assume that system (\ref{gen}) supports the
single-phase periodic travelling wave solutions for $\phi=(\rho,
u)$:
\begin{equation} \label{trav}
\phi (x,t)= \phi(x-ct) \, , \qquad
\phi(\theta+2\pi/k)=\phi(\theta) \ ,
\end{equation}
which are parametrised by four constants, say $\bar \rho, \bar u,
k, a$, where $\bar \rho$ and $\bar u$ are the mean density and
mean velocity respectively, and $k$ and $a$ are, as usual, the
wavenumber and the amplitude. We assume that the potential curve
$G(\lambda)$ in the travelling wave equation
$\lambda_{\theta}^2=-G(\lambda)$ has generic properties outlined
in Section 4 (see (\ref{quadr}), (\ref{Gu})) such that the
solution (\ref{trav}) admits the limiting transitions to a linear
wave as $a \to 0$ and  to a solitary wave as $k \to 0$.

 The described class of systems (\ref{gen})  is quite broad and includes some
known integrable models such as defocusing nonlinear Schr\"odinger
equation and Kaup-Boussinesq system \cite{kaup76}, \cite{egp01}.
As physically important examples of bi-directional non-integrable
systems that possess the above  general properties (including the
gas dynamics form of the dispersionless limit) one can indicate
the Green-Naghdi system for fully nonlinear shallow water gravity
waves \cite{gn76} and its multi-layer generalisations \cite{cc99},
the systems for nonlinear ion-acoustic and magnetoacoustic waves
in collisionless plasma \cite{karp75}, \cite{wh65},  and many
others.

We consider  initial data  for the system (\ref{gen})  in the form
of a step for the variables $\rho$ and $u$:
\begin{equation} \label{decay}
t=0: \ \ \rho=\rho^-, \  \   u=u^- \  \hbox{for} \  x <0;  \quad
\rho=\rho^+, \      u=u^+\,   \hbox{for}  \  x>0 \, ,
\end{equation}
where $\rho^{\pm}$ and $u^{\pm}$ are constants.

Analytical studies of the decay of an initial discontinuity
problem in  integrable dispersive wave equations (see for instance
\cite{eggk95},  \cite{kod99},  \cite{egp01}) as well as direct
numerical simulations for non-integrable systems ( see \cite{gm84}
\cite{hl91}, \cite{llv93} and references therein) suggest that the
asymptotic solution for the decay of an arbitrary initial
discontinuity problem generally consists of three constant states
separated by two expanding waves: centred rarefaction wave(s)
and/or dispersive shock(s), which is quite natural  taking into
account the ``two-wave'' nature of the system (\ref{gen}).  The
structure and the qualitative properties of the dispersive shock
are the same as in the case of simple-wave lead dispersive
equations  (see Section 4) so we will model such a dispersive
shock by the expansion fan solution of the corresponding Whitham
system, which can be represented in the form
\begin{equation}\label{avgen}
\frac{\partial } {\partial t}\overline{P}_j(\bar \rho, \bar u, k,
a) +\frac{\partial }{\partial x}\overline{Q}_j(\bar \rho, \bar u,
k, a)=0\, , \quad j=1,2,3 \, ,
\end{equation}
\begin{equation} \label{94}
 \frac{\partial}{\partial t} k+
\frac{\partial }{\partial x}\omega (\bar \rho, \bar u, k, a)=0\, ,
\end{equation}
where $\bar \rho$ and $\bar u$ are the density and  velocity
averaged over the family (\ref{trav}), $k$ is the wavenumber and
$a$ is the wave amplitude; also as follows from the formulation
$\overline{P_1}=\bar \rho$, $\overline{P_2} = \overline{\rho u}$.
Using the arguments presented in Section 3.1, we {\it postulate}
the following fundamental property of the Whitham equations
(\ref{avgen}) for weakly dispersive nonlinear wave equations
(\ref{gen}):

{\bf  The averaged ``dispersive-hydrodynamic" conservation laws
(\ref{avgen}) admit exact reductions to the dispersionless limit
system (\ref{euler}) for $a=0$ or $k=0$:}
\begin{equation} \label{euler1}
  \bar \rho_t +
 (  \bar \rho \bar u)_x = 0 \, , \quad (\bar \rho \bar u)_t +
 (\bar \rho \bar u^2 + p(\bar \rho))_x = 0 ,
\end{equation}
so that the ``energy'' equation for $\bar P_3$ becomes a
consequence of (\ref{euler1}). Of course, in concrete instances
this physically transparent property can be easily verified by
direct calculation using the asymptotic behaviour of the potential
curve $G(\lambda)$ for the nearly linear and nearly soliton
configurations (see (\ref{quadr}), (\ref{Gu})).

The wave number conservation law (\ref{94}) in the linear limit
assumes the form (cf. (\ref{wh0n})):
\begin{equation}\label{lcw}
a=0, \, \qquad   k_t+ (\omega_0 (\bar \rho, \bar u, k))_x=0\ \,
\end{equation}
where $\omega =\omega_0 (\bar \rho, \bar u, k)$ is the linear
dispersion relation for the system (\ref{gen}) obtained by
considering  an infinitesimally small perturbation of the ground
state $\rho=\bar \rho, \ u=\bar u$. Owing to the two-wave nature
of the system (\ref{gen}) there are two branches of the linear
dispersion relation. To avoid unnecessary complication, it will be
assumed  that, unless otherwise specified, we consider the branch
corresponding to the right-propagating waves.

Now we can formulate the natural boundary conditions of the
Gurevich-Pitaevskii type for a bi-directional Whitham system
(\ref{avgen}), (\ref{94}). Again, for simplicity of  presentation
we assume the negative dispersion wave pattern and later will
remove this restriction in the formulation of the transition
conditions. In the decay of an initial discontinuity problem we
are interested in the similarity solutions of the modulation
equations so we present the natural matching problem for
modulation variables at once in the form analogous to (\ref{bc2})
\begin{equation}
\begin{array}{l}
x=s^-t : \qquad a =0\, , \ \ \bar \rho = \rho^-\, , \ \ \bar u =u^- \, , \\
x=s^+t : \qquad k=0\, , \ \ \bar \rho = \rho^+ \, , \ \ \bar u=u^+
\, ,
\end{array}
\label{bc6}
\end{equation}
where the dependencies of the edge speeds $s^{\pm}$ on the initial
jump parameters $\rho^+, \rho^-, u^+, u^-$ are to be found.

 It is clear that the similarity solution of the
fourth order Whitham system is parametrised  by only three
constants. Therefore, one must impose an additional restriction on
the values of the parameters $\rho^{\pm}, u^{\pm}$ to select the
family of admissible jumps across the dispersive shock. Of course,
the necessity of an additional jump condition for the dispersive
shock generated in the decay of an initial discontinuity is
inherently implied by the two-wave nature of the system under
consideration. Finding this restriction is equivalent to the
extracting the family of the initial discontinuities resolving
into  a single dispersive shock propagating in a given direction
(we associate the direction of the dispersive shock propagation
with the direction of the corresponding linear characteristic). In
dissipative gas dynamics, such admissible discontinuities  are
selected by the Rankine-Hugoniot conditions following from the
balance of mass, momentum and total energy across the shock. So,
before we proceed with the determination of the dispersive shock
edges as we did in the simple-wave led equations, our task is to
obtain a replacement for the classical shock curve in the
bi-directional dispersive hydrodynamics which would have a form of
a relationship
\begin{equation} \label{3p}
\Phi(\rho^+, \rho^-, u^+, u^-)=0 \,
\end{equation}
We note that in the conservative dispersive shock transition, the
pressure $p$ is the function of the density only (the
thermodynamic entropy does not change in the course of the
dispersive shock propagation) so, in contrast to the classical
shock theory \cite{lanlif87}, the pressure should not appear in
the transition relation (\ref{3p}).
\subsection{Dispersive shock curve and ``entropy'' conditions}

We proceed with the dispersive shock curve using a simple method
recently proposed in  \cite{ekt05}. The idea is, using the
conservative nature of the dissipationless dispersive shocks to
``reverse'' the dispersive shock solution in time. It is then
``natural'' to expect that the dispersive shock would convert into
a rarefaction wave of the dispersionless Euler system
(\ref{euler}). One can then ``extract'' the required transition
relation from  simple analytic solution for this rarefaction wave
which is easily constructed using Riemann invariants (see for
instance \cite{lanlif87}, \cite{Wh74}). Of course, such an
intuitive argument requires mathematical justification since it is
not obvious why the ``inverse'' resolution should happen through a
single rarefaction wave.

Our construction  is based on the following geometric
consideration in the characteristic $(x,t)$-plane. First we
represent the Euler system (\ref{euler}) in the Riemann invariant
form
\begin{equation} \label{rim}
r_t  + V_r(r,l)r_x=0 \, , \qquad l_t  + V_l(r, l)l_x=0 \, ,
\end{equation}
where
\begin{equation} \label{rV}
r=u +  \int ^{\rho}_{\rho_0}\frac{\sigma(\rho')}{\rho'}d\rho' \, ,
\quad l=u -  \int
^{\rho}_{\rho_0}\frac{\sigma(\rho')}{\rho'}d\rho'\, ; \quad
V_{r,l}=u \pm \sigma(\rho)\, .
\end{equation}
Here $\rho_0=const$ and $\sigma(\rho)=(dp/d\rho)^{1/2}$ is the
``sound speed''. The initial conditions (\ref{decay}) are
rewritten in terms of the Riemann invariants as
\begin{equation}
\label{decay4} r=r^-\, , \ l=l^- \ \  \hbox{for} \ x<0 \, ; \qquad
r=r^+\, , \ l=l^+ \ \  \hbox{for} \ x>0 \, ,
\end{equation}
where $r^{\pm}=r(\rho^{\pm}, u^{\pm})$, $l^{\pm}=l(\rho^{\pm},
u^{\pm})$. The corresponding four characteristic directions at
$t=0$ are $V_r^{\pm}\equiv V_r(r^{\pm}, l^{\pm})$ and
$V_l^{\pm}\equiv V_l(r^{\pm}, l^{\pm})$.

We consider the ``right-propagating'' dispersive shock, which
occurs when the characteristics of the $r$-family intersect i.e.
when $V_r^->V_r^+$. Then, assuming the modulation description of
the dispersive shock with the aid of the expansion fan solution of
the Whitham equations, one can observe that, due to (assumed)
hyperbolicity the solution of the GP problem (\ref{avgen}),
(\ref{94}), (\ref{bc6}) can be, in principle, constructed
geometrically using characteristics. The continuity matching
(\ref{bc6}) for the mean values  in such a construction is
replaced with the equivalent continuity matching for the Whitham
and Euler characteristics  (see \cite{te99}). One can also observe
that although such a construction can be very complicated for
$t>0$, it can be readily realised for negative $t$. Indeed, since
the Whitham expansion fan  has zero width at $t=0$ (hyperbolicity)
the initial conditions (\ref{decay4}) are specified entirely in
the external, ``dispersionless'' domain of the space-time of the
GP problem. Thus the global solution of the GP problem can be
continued backwards along the  characteristics of the $2 \times 2$
dispersionless limit equations (\ref{rim}). By construction, this
solution must (i) be three-parametric (see (\ref{3p})) and (ii)
satisfy the inequality $V_r^->V_r^+$ at $t=0$.

There is a unique three-parametric solution to the hyperbolic
 system (\ref{rim}) in the lower $(xt)$ half-plane
satisfying the described restrictions.  This solution is a centred
expansion fan (in $(-x, -t)$ - coordinates) given by the
expressions
\begin{equation} \label{const1} l=l_0=constant
\end{equation}
\bea
\   \  \ r&=& r^{-}; \qquad  \qquad \qquad x > a_2t ;\nonumber \\
V_r(r, l_0) &=& x/t; \    \qquad  \qquad a_1t \le x \le a_2t; \label{decaysol1}\\
 \   \  \ r&=& r^{+}; \qquad  \qquad \qquad x <a_1 t . \nonumber
\eea Here
\begin{equation} \label{a} a_{1}=V_r^{+}\, , \ \ a_{2}=V_r^{-}\, ,
\qquad a_2>a_1 \, .
\end{equation}
The solution (\ref{const1}) -- (\ref{a}) is characterised by three
parameters $(r^+, r^-, l_0)$ and {\it it exists for all $t<0$}.
Since this solution  represents a continuation, along the
characteristics of the {\it full} solution to the  GP problem, one
can regard this solution evaluated at a fixed moment $t_0<0$ as
new initial conditions for the same GP problem. Therefore, one can
extract from it the global restriction (\ref{3p}) imposed on
possible values of the hydrodynamic variables at the opposite
sides of the dispersive shock. This restriction follows from
(\ref{const1}) and has the form
\begin{equation} \label{1}
l^-=l^+ \, ,
\end{equation}
 which  in view of the relationships (\ref{rV}) can be represented
in an explicit gas-dynamic form,
 \begin{equation} \label{GM}
 u^--u^+=\int_{\rho^+}^{\rho^-}\frac{\sigma(\rho)}{ \rho}d\rho \, .
 \end{equation}
 Given the state in front of the dispersive shock $(\rho^+, u^+)$
this relation yields all admissible states
 $(\rho^-, u^-)$ behind it i.e. it represents the equation of the
 $\rho$-$u$ diagram of the dispersive shock.
One can see that the whole above construction is subject to the
additional inequalities
 \begin{equation} \label{in2}
V_{l}^{-} <s^-<V_{r}^{-}\, , \quad   V_{r}^{+} <s^+\, , \quad
s^+>s^-\, .
\end{equation}
 These inequalities ensure fulfillment of our original requirement
about the single-wave resolution of a step (i.e. three-parametric
 solution). Indeed, the number of
parameters characterising the solution of the hyperbolic system in
some domain is equal to the number of families of characteristics
transferring given initial or boundary data {\it into} this domain
(see for instance \cite{lanlif87}). The inequalities (\ref{in2})
require that only three of the four families of classical
characteristics $x/t=V_{l,r}^{\pm}$ (namely, $x/t=V_{r}^{-}$ ,
$x/t=V_{r}^{+}$, and $x/t=V_{l}^{+}$) transfer initial data
(\ref{decay4}) from the $x$-axis into the dispersive shock domain
(see Fig.4). One can see that the inequalities (\ref{in2})
represent an extension of the ``entropy'' conditions (\ref{in0})
to the bi-directional case.
\begin{figure}[ht]
\centerline{\includegraphics[width=7cm,height=
13cm,clip]{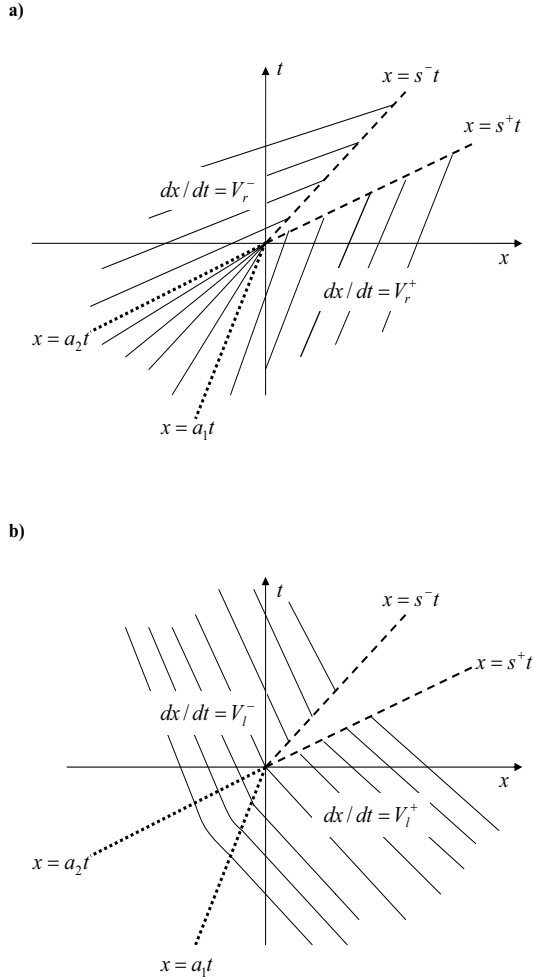}} \vspace{0.3 true cm} \caption{Qualitative
 behaviour of characteristics in the ``simple-wave'' decay in
 dispersive hydrodynamics. Broken lines:
the dispersive shock boundaries. Dotted lines: the ``mirror''
expansion fan boundaries. (a) Families $dx/dt=V_r^{\pm}$ transfers
values of $r$, (b) Families $dx/dt=V_l^{\pm}$ transfers values of
$l$.} \label{fig4}
\end{figure}

Now we  discuss briefly the meaning of the obtained transition
relation (\ref{GM}). One can see that it coincides with the
relationship between any two pairs  $\rho$, $u$ in the simple wave
solution of the isentropic gas dynamics (see for instance
\cite{lanlif87}). A nontrivial fact is that the corresponding wave
is the simple wave of {\it compression} which breaks after a
certain time interval. Contrastingly, the similarity solution of
the Whitham equations satisfying the relationship (\ref{const1})
and describing the expanding dispersive shock does not break. This
solution essentially represents a {\it compression fan}, which
does not exist in classical gas dynamics. It is natural to call
the dispersive shock satisfying the relationship (\ref{const1})
{\it a simple dispersive shock}. We emphasize that, according to
the matching conditions (\ref{bc6}) the relationship (\ref{GM}) is
only valid {\it for the boundary values} of the modulation
parameters $\bar \rho$ and $\bar u$ and, of course, does not hold
within the dispersive shock region.

In the  solutions of the GP problem for integrable systems, the
simple dispersive shock condition  is a mere consequence of the
constancy of one of the Riemann invariants of the Whitham system
(see \cite{eggk95}, \cite{egp01} for instance). In non-integrable
case, when the Riemann invariants are not available for the
Whitham system, the condition (\ref{GM}) is not obvious at all.
Also, one should keep in mind that inequalities (\ref{in2})
represent a necessary part of the transition conditions.

We recall that  the above consideration  was concerned with the
dispersive shocks propagating to the right. For the
left-propagating dispersive shocks an analogous system of
relations would include the  zero jump condition for the classical
Riemann invariant $r$  instead of $l$ in (\ref{const1}) and the
``entropy''  inequalities analogous to (\ref{in2}) would have the
form
\begin{equation}\label{in3}
V_l^+<s^+<V_r^+ \,,  \quad s^-<V_l^-\, , \quad s^+>s^-\,.
\end{equation}
The simple dispersive shock curve in the form (\ref{GM}) has been
proposed for the first time, using intuitive physical arguments,
by Gurevich and Meshcherkin \cite{gm84} in the context of
collisionless shocks in plasma. Later, this condition has been
interpreted in \cite{te99} in terms of the ``local Riemann
invariant'' transport through the Whitham zone.

We note in conclusion that, in a more general case, when the
dispersionless limit has the form other than isentropic gas
dynamics, one should use the simple dispersive shock curve in the
invariant form (\ref{1}).

\subsection{Speeds of the dispersive shock edges}
We now proceed with the dispersive shock edges similarly to
Sections 3.2, 3.3. We consider a characteristic equation for the
Whitham system (\ref{avgen}), (\ref{94})
\begin{equation}\label{char3}
b_1^{(i)}d_i \bar \rho+ b_2^{(i)}d_i \bar u + b_3^{(i)}d_i k+
b_4^{(i)}d_i a=0 \, ,
\end{equation}
where ${\bf b^{(i)}}(\bar \rho, \bar u, k, a)$ is the left
eigenvector of the coefficient matrix $A(\bar \rho, \bar u, k, a)$
of the modulation system (\ref{avgen}), (\ref{94}) and $i$ is a
number of the characteristic. Let $a=0$ on this characteristic.
Then for the remaining three variables we have the ordinary
differential equation
\begin{equation}\label{3form}
b_1^{(i)}(\bar \rho, \bar u, k, 0)d_i \bar \rho+ b_2^{(i)}(\bar
\rho, \bar u, k, 0)d_i \bar u + b_3^{(i)}(\bar \rho, \bar u, k,
0)d_i k=0
\end{equation}
which necessarily is a characteristic equation for the  reduced as
$a=0$ modulation system (\ref{avgen}), (\ref{94}):
\begin{equation} \label{red}
 a=0, \quad   \qquad \bar \rho_t +
 (  \bar \rho \bar u)_x = 0 \, , \quad (\bar \rho \bar u)_t +
 (\bar \rho \bar u^2 + p(\bar \rho))_x = 0 , \quad k_t+
 ( \omega_0(\bar \rho, \bar u, k))_x=0\, .
\end{equation}
Now we impose an additional constraint
\begin{equation}\label{constr}
F(\bar \rho, \bar u)=C_0 \, ,
\end{equation}
where $C_0$ is a constant. It will be shown later that this
constraint leads to a self-consistent set of the transition
conditions. One can see that Eq.~(\ref{constr}) is consistent with
the reduction (\ref{red}) and, therefore, is compatible with
(\ref{3form}) iff the function $F(\bar \rho, \bar u)$ is one of
the Riemann invariants $l,r$ defined by (\ref{rV}).
 Setting Eq.~(\ref{constr}) into Eq.~(\ref{3form}) we obtain a $2$-form
 which can be integrated
 to give the local  relationship between admissible values of $\bar \rho$
and $k$ {\it on the characteristic},
\begin{equation}\label{int1}
\Phi_i(\bar u, k;C_0)=C_1\, ,
\end{equation}
$C_1$ being a  constant of integration.

Alternatively, the relationship (\ref{int1}) can be obtained by a
direct substitution of Eq.~(\ref{constr}) into the system
(\ref{red}) and then, by looking for the solution in the form
$k(\bar u)$. Substitution of Eq.~(\ref{constr}) into (\ref{red})
yields (cf. (\ref{wh0n})):
\begin{equation}\label{wh3}
a=0, \qquad \bar u_t+V(\bar u) \bar u _x=0\, , \qquad k_t+
(\omega_0(\bar u, k))_x=0 \, ,
\end{equation}
where (see Eq.~(\ref{rV}))
\begin{equation}\label{rel3}
V(\bar u) \equiv  V_r(\rho(\bar u), \bar u)\, , \qquad
\omega_0(\bar u, k) \equiv \omega_0(\bar \rho(\bar u), \bar u, k)
\, .
\end{equation}
Thus, the problem essentially reduces to the simple-wave led case
considered in Sections 3, 4 and, therefore, yields the ordinary
differential equation (\ref{ode1}) for $k(\bar u)$. In our case,
however, this equation is additionally parametrised by the
unidentified (yet) function $F(\bar\rho, \bar u)$ and a
 constant $C_0$
 in (\ref{constr}) defining the dependence $\bar \rho (\bar u)$ in
 (\ref{rel3}). These can be found by applying the matching
 conditions (\ref{bc6}) considered in the four-dimensional space of
 the field variables $\bar \rho, \bar u, k, a$.
 Since the integral (\ref{constr}) does not contain $a$ and $k$ it
 must apply to both boundaries $a=0$ and $k=0$, which requires $F(\rho^-, u^-)=F(\rho^+,
 u^+)$. Then, to be consistent with the dispersive shock curve (\ref{1}) one
must identify $F(\bar \rho, \bar u) \equiv l(\bar \rho, \bar
 u)$, which immediately yields $C_0=l^-= l^+$.
Now, the trailing edge speed $s^-$ is
 found with the aid of Eq. (\ref{spm}a).

The leading edge edge is handled in exactly the same way, by
reducing the ``hydrodynamic'' part of  the modulation system for
$k=0$ to the simple wave equation with the aid of the constraint
(\ref{constr}). Then one proceeds with the conjugate variables as
in Section 3.3 to get a general simple-wave analog of the ordinary
differential equation (\ref{lkdv0}) i.e. Eq. (\ref{ode2}) for
$\tilde k (\bar u)$.  At last, using the dispersive shock curve
(\ref{1}) the function $F(\bar \rho, \bar u)$ in (\ref{rel3}) is
identified with the same classical Riemann invariant $l$, which
justifies  self-consistency of the whole construction. Then the
leading edge speed is found via the conjugate wavenumber $\tilde
k$ by Eqs.~(\ref{spm}b), (\ref{ode2}) where all the necessary
ingredients are given by Eq.~(\ref{rel3}).

The whole construction is subject to the ``entropy conditions''
(\ref{ent}) which should be complemented by an additional
inequality $ s^->V_{l}^{-}$ (see (\ref{in2})) occurring due to the
presence of the fourth characteristic family in the bi-directional
problems (see \cite{ekt05} for an example where this inequality is
violated).

It is clear that analogous formulas can be obtained for the left-
propagating dispersive shock: one should just replace  in
Eq.~(\ref{rel3}) $V_r(\bar u)$ with $V_l(\bar u)$ and $l(\bar
\rho, \bar u)=C_0$ with $r(\bar \rho, \bar u)=C_0^*$ where
$C_0^*=r^-=r^+$ according to the corresponding dispersive shock
curve. Also, one should use the ``left'' branch of the linear
dispersion relation and the ``entropy'' conditions (\ref{in3}).

Some additional details of the derivation of the trailing and the
leading edge curves for bi-directional systems can be found in
\cite{ekt05}.

\subsection{Example: fully nonlinear ion-acoustic collisionless shock in plasma}
As an
example of an effective construction of the dispersive shock
conditions in a non-integrable system we consider classical system
of equations describing finite-amplitude ion-acoustic waves in a
two-temperature ($T_{e}\gg T_{i}$) collisionless plasma (see for
instance \cite{karp75})
\begin{eqnarray}
&&\rho_t + (\rho {u})_x = 0 ,\nonumber \\
& &u_t + uu_x + \varphi_x = 0 , \label{ion}\\
&&\varphi_{xx} =e^{\varphi}-\rho \, . \nonumber
\end{eqnarray}
Here $\rho $ and $u$ are  the ion density and velocity and
$\varphi$ is the electric potential; all dependent variables are
dimensionless. We note that the system does not contain the time
derivative of $\varphi$ so $\varphi$ is not a ``real'' dependent
variable as concerns the $2 \times 2$ representation (\ref{gen}).
A direct numerical simulation of the decay of an initial
discontinuity in Eqs.~(\ref{ion}) has been performed in
\cite{gm84}.

The system (\ref{ion}) supports periodic travelling waves allowing
for linear and solitary wave limits and also possesses (at least)
four conservation laws \cite{gke90}. Thus we can apply the methods
developed in this paper for obtaining the dispersive shock
conditions. We consider the right-propagating simple dispersive
shock.

In the dispersionless limit $\varphi=\ln \rho$, which yields the
Euler isentropic gas-dynamic equations (\ref{euler}) with the
equation of state  $p(\rho)=\rho$. The corresponding  Riemann
invariants and characteristic velocities are (see (\ref{rim}),
(\ref{rV})):
\begin{equation}\label{rVis}
l=u-\ln\rho\, ,\quad  r=u+\ln \rho\, , \quad V_l=u-1\,, \quad
V_r=u+1 \, ,
\end{equation}
while the linear dispersion relation for the right-propagating
modulated waves has the form
\begin{equation} \label{dis}
\omega_0(k, \bar \rho, \bar u)=k[\bar u+(1+{k^2}/{\bar
\rho})^{-1/2} ]\, .
\end{equation}
The dispersive shock transition curve (\ref{GM})  then assumes the
form $u^--u^+=\ln (\rho^-/ \rho^+)$. Without loss of generality we
put $u^+=0$, $\rho^+=1$. Then the relationship  between $\bar
\rho$ and $\bar u$ in Eq.~(\ref{rel3}) becomes $\bar u= \ln \bar
\rho$. As a result, we get all the necessary ingredients  for the
basic ordinary differential equations (\ref{ode1}), (\ref{ode2}):

\begin{equation} \label{ior}
V(\bar u) =\bar u +1 \, , \qquad \omega_0(\bar u, k)= k [ \bar u +
(1+{k^2}/{e^{\bar u}})^{-1/2} ] \, .
\end{equation}

 Then the equation (\ref{ode1})  after elementary
transformations, assumes the form with separated variables
\begin{equation} \label{eq} \frac{d \alpha}{d \bar u}=
-\frac{(1+\alpha)^2\alpha}{2(1+\alpha+\alpha^2)} \, , \qquad
\alpha(1)=1 \, ,
\end{equation}
where $ \alpha=(1+k^2/e^{\bar u})^{-1/2}$. Integrating (\ref{eq})
we get
\begin{equation} \label{integral} \bar
u+2\ln\alpha+\frac{1-\alpha}{1+\alpha}=0 \, .
\end{equation}
Now, using the formula (\ref{spm}a) we obtain a simple equation
determining velocity of the trailing edge $s^-$ in terms of the
density ratio across the dispersive shock $d =\rho^-/\rho^+ =
\rho^-$,
\begin{equation} \label{c2is} \ln d +\frac{2}{3}\ln(s^--\ln d )=
\frac{(s^--\ln d)^{1/3}-1}{(s^--\ln d)^{1/3}+1} \, .
\end{equation}
The leading edge is handled in a completely analogous way. The
solitary wave dispersion relation  is obtained from the linear
dispersion relation (\ref{ior}) as $\tilde \omega_s(\bar u, \tilde
k )=-i \omega_0(\bar u, i \tilde k)$ which is to say
\begin{equation} \label{}
 \qquad \tilde \omega_s(\bar u, \tilde k)= \tilde k [ \bar u +
(1-{\tilde k^2}/{e^{\bar u}})^{-1/2}] \, .
\end{equation} Then,
integrating (\ref{ode2}) we obtain $\tilde k(\bar u)$ (it is also
convenient to introduce $ \tilde \alpha=(1-\tilde k^2/e^{\bar
u})^{-1/2}$ as an intermediate variable  instead of $\tilde k$).
Setting it into (\ref{spm}b) we eventually get the equation for
the leading edge
\begin{equation} \label{c1is}
2\ln s^+ - \frac{s^+-1}{s^++1} =\ln d \, .
\end{equation}

\begin{figure}[ht]
\centerline{\includegraphics[width=7cm,height=
7cm,clip]{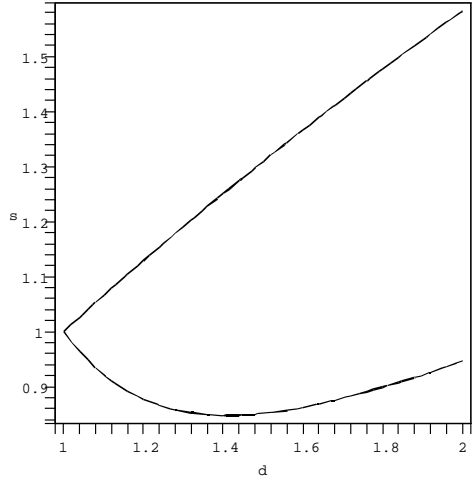}} \vspace{0.3 true cm} \caption{Speeds of the
edges of the ion-acoustic dispersive shock transition versus
density ratio $d=\rho^-/\rho^+$ across the shock; upper graph:
$s^+$-- leading edge, lower graph: $s^-$ -- trailing edge}
\label{fig5}
\end{figure}
It can be easily  verified that the ``entropy'' conditions
(\ref{in2}) are satisfied for all values of the density ratio
across the dispersive shock. The curves $s^{+} (d)$ and $s^{-}
(d)$ are presented in Fig.5 and demonstrate very good agreement
with the results of direct numerical simulation of the decay of an
initial discontinuity for the system (\ref{ion}) presented in
\cite{gm84} (see Fig. 6 in this paper analogous to our Fig.5).
From theoretical point of view such an agreement can be regarded
as a strong indication of validity of the modulation theory in a
certain class of non-integrable initial value problems where
rigorous derivation of the Whitham asymptotic as a zero-dispersion
limit is not available.

The weakly nonlinear asymptotic decompositions of (\ref{c2is}) and
(\ref{c1is}) for small $ \delta = d - 1 \ll 1$, have the form
\begin{equation} \label{kdvis}
s^-\approx 1-\delta\, , \qquad  s^+ \approx1+\frac{2}{3}\delta\, ,
\end{equation}
an, as a matter of fact, agree with the Gurevich-Pitaevskii
solution for the KdV solution (\ref{spm0}) which is another
confirmation of  validity of the obtained modulation solution.
Curiously, as is clearly seen from Fig. 5 , the fully nonlinear
dynamics of the leading (solitary wave) edge $s^+$ is very well
approximated by the weakly nonlinear asymptotic (\ref{kdvis}) in a
broad range of the density ratios $d$ while the speed of the
trailing (harmonic) edge $s^-$ demonstrates significant
qualitative and quantitative deviations from its weakly nonlinear
counterpart even for quite moderate values of $d$.

A detailed comparison of the modulation transition conditions
obtained here with results of direct numerical simulation for the
(non-integrable) Green -- Naghdi system describing fully nonlinear
shallow-water waves will be published in \cite{egs}.
\section{Some  restrictions}
We first address the accuracy of the obtained results in the
context of the original (dispersive) system and outline some
restrictions of our analysis of the dispersive shocks. Our main
assumption  was about modelling the a dispersive shock with the
aid of the expansion fan solution of the Whitham equations. The
Whitham equations, being obtained by an asymptotic procedure, have
inherent accuracy restrictions on their applicability to certain
wave regimes. In our (GP) formulation, in contrast to the formal
zero-dispersion limit approach, the small dispersion parameter
determining the accuracy of the Whitham approximation is assumed
to occur {\it in the solutions} and is defined by the ratio of the
typical wavelength to the characteristic scale of the modulation
variations (in our case, the width of the dispersive shock $L \sim
t$), i.e.
\begin{equation}\label{acc1}
\epsilon \sim (kt)^{-1} \ll 1 \, .
\end{equation}
which implies that that modulation description of the dispersive
shock is valid asymptotically as $t \gg 1$. The Whitham method of
averaging conservation laws \cite{wh65} also requires that the
relative variations of modulation parameters over the
characteristic wavelength scale $\sim k^{-1}$ be $\mathcal{O}
(\epsilon$) which yields the criterion
\begin{equation}\label{acc2}
|k_x| k^{-1} \sim t^{-1} \, .
\end{equation}
One can now observe that, in the modulation solution for the
dispersive shock, for {\it any} $t$, however large, there is a
certain vicinity of the of the leading edge where both criteria
(\ref{acc1}), (\ref{acc2}) are violated due to small (and rapidly
changing) values of $k$. This is basically a reflection of
non-uniformity occurring as $k \to 0$ in the formal perturbation
decompositions (see e.g. \cite{kuzmak}) equivalent to leading
order to the Whitham method. So a natural question of
applicability of the modulation description to the dynamics of the
leading edge of the dispersive shock arises. While this question
does not appear in the the rigorous zero-dispersion limit approach
\cite {laxlev83} in which the Whitham equations are {\it derived}
as a certain asymptotics in the initial-value problem and
$\epsilon$ does not depend on $t$, it has to be addressed in our
case when the Whitham description is {\it assumed} on heuristic
foundations in the absence of rich integrable structure.

We first determine the  behaviour of the function $k(x)$ in the
vicinity of the leading edge of the simple dispersive shock for
fixed $t$. It is instructive  to make first an estimate for the
KdV case, which can be made by using the wave number conservation
law in the self-similar form $-sk'+\omega'=0$, $s=x/t$. Now
setting the asymptotic decompositions for $k$ and $\omega$ as $m'
\ll 1$,  which are readily obtained from (\ref{L}) we get for the
vicinity of the leading edge $s^+-s \ll 1$ the following
asymptotic behaviour: $k \sim k_0/\ln\delta^{-1}$, $\omega- k s^+
\sim k \delta$, $\delta \sim (s^+ - s)(\ln1/(s^+-s))^{-1}$. It is
clear that this behaviour is automatically generalised to other
nonlinear dispersive systems for which the potential curve $G(u)$
in the travelling waves has the asymptotic behaviour
(\ref{quadr}), ({\ref{Gu}) in the nearly linear and nearly soliton
configurations.

Now, using the obtained asymptotics for $k$, we estimate the
relative widths $\sigma=\Delta x /L$ of the vicinities of the
leading edge $x^+=s^+ t$  where the criteria (\ref{acc1}),
(\ref{acc2}) are violated. For (\ref{acc1}) we get $\sigma_1 \sim
\exp(-t)$ while for (\ref{acc2}) we get $\sigma_2 \sim t^{-1}\ln t
$, i.e $\sigma_2 \gg \sigma_1$. Still, $\sigma_2(t) \to 0$ as $t
\to \infty$
 and, therefore, the Whitham description  of the dispersive
shock is asymptotically as $t \gg 1$ valid for all $s^-t<x<s^+t$.

The modulation approach used in this paper requires the existence
of the single-phase periodic solutions. For non-integrable systems
such solutions often exist only within a certain domain of
parameters. Typically the role of the ``critical'' parameter is
played by the wave amplitude, so that for $a>a_{cr}$ the periodic
solution (or a solitary wave) ceases to exist because of
occurrence of breaking or cusp-type singularities.  This usually
violates the ``single-flow'' type assumptions used in the
derivation of the original system. For instance, for ion-acoustic
system (\ref{ion})  the critical solitary wave amplitude for the
potential $\varphi$ is $a_{cr} \approx 1.3$ \cite{sagdeev}. For
$a_s>a_{cr}$ the ``hydrodynamic'' single-flow description of the
two-temperature plasma becomes not applicable and a more general,
kinetic plasma theory should be used. The critical value for the
density ratio across the ion-acoustic dispersive shock
$\Delta_{cr}$ can be found by setting  the critical value of the
solitary wave speed $c_s(a_{cr}) \approx 1.6$ \cite{gm84} into
Eq.~(\ref{c1is}): $s^+(\Delta_{cr})\approx 1.6$, which yields
$\Delta_{cr} \approx 2.0$. Another possible restriction on the
application of the obtained transition conditions partially
addressed in Section 4 is imposed by the requirement of the
modulational stability of the dispersive shock solution. For some
equations the type of the modulation system can change from
hyperbolic to elliptic depending on the initial data. One of the
 examples is provided by the so-called Kaup-Boussinesq
(integrable) system describing bi-directional shallow water waves
where the linear dispersion relation yields complex frequencies if
the wavenumber exceeds a certain value \cite{egp01}. The
corresponding restrictions on the jump values across the
dispersive shock can be found from the conditions (\ref{stab}).

\section{Conclusions}
A new method has been proposed to analyse the dispersive shock
transition for a broad class of weakly dispersive nonlinear wave
equations. No assumptions of integrability have been made which,
in particular, allows one to apply the developed method to
problems of fully nonlinear dispersive wave dynamics. The
transition conditions have been derived by assuming the
description of the dispersive shock with the aid of the expansion
fan solutions of the associated modulation (Whitham) equations.
The Whitham system was assumed to be hyperbolic for the solutions
of our interest, which seems a reasonable assumption for the
outlined class dispersive systems (at least for a certain type of
initial data). This assumption can also be viewed as an inference
from direct numerical simulations.

The analysis has been performed using the generalisation of the
so-called Gurevich-Pitaevskii problem formulated originally for
the averaged KdV equation. It has been shown that the
Gurevich-Pitaevskii type natural boundary conditions for the mean
flow in the dispersive shock region can be translated into
information about degeneration of the families of characteristics
along the dispersive shock boundaries. The latter implies certain
restrictions on admissible values of the hydrodynamic and wave
parameters at these boundaries. It has been shown that, for the
problem of an initial step resolution, these restrictions can be
found without presence of the Riemann invariant structure for the
Whitham system.

As a result, we have derived a complete set of transition
conditions for dispersive shocks linking two different constant
hydrodynamic flows. The analysis has been performed as for
single-wave equations so for bi-directional systems. The
transition conditions have been derived in a general form and
include: i) a ``dispersive" analog of the traditional jump
conditions; ii) the equations for the boundary curves of the
dispersive shock; iii) a set of inequalities similar to the
entropy conditions of traditional gas dynamics.

Remarkably, the whole set of the transition conditions is
constructed in terms of the linear dispersion relation and the
dispersionless nonlinear characteristic velocities of the system
under study. The obtained conditions allow one to ``fit'' a
 dispersive shock into the classical dispersionless solution without complicated analysis of
 the internal structure of the dispersive shock similarly to classical shock theory
in traditional gas dynamics. The method can be useful not only for
the analysis of non-integrable dispersive
 systems where full solution is not available but also for integrable systems
 when one is interested only in main physical parameters of the dispersive shock transition.
 As an additional bonus, the developed method allows for obtaining the lead solitary wave amplitude,
 the major parameter observed in  experiments.

 \vspace{0.5cm} \noindent {\bf Acknowledgements}

\noindent I would like to thank my Mathematics teacher and friend
Alexander Krylov and my life teacher Alexander Gurevich for
introducing me to this exciting topic. I thank Vadim Khodorovskii
and Alexandra Tyurina, my co-authors in several papers where a
part of the results presented here has been published. I am
grateful to Stephanos Venakides for valuable comments concerning
general meaning of the obtained results, to  Jacques Vanneste for
asking some important questions, and to Roger Grimshaw and Anatoly
Kamchatnov for their interest in this work and useful discussions.
Special thanks to Andrew El for inspiring remarks during the work
on this paper.

\end{document}